\documentclass{ar-1col-S2O}
\usepackage[numbers]{natbib}
\usepackage{url}
\newcommand{\mll}         {\ensuremath{m_{\rm ll}}}
\newcommand{\mee}         {\ensuremath{m_{\rm ee}}}
\newcommand{\ptee}         {\ensuremath{p_{\rm T,ee}}}
\newcommand{\ptll}         {\ensuremath{p_{\rm T,ll}}}
\newcommand{\vtwoll}         {\ensuremath{v_{\rm 2,ll}}}

\newcommand{\snn}          {\ensuremath{\sqrt{s_{\mathrm{NN}}}}}
\newcommand{\pt}           {\ensuremath{p_{\rm T}}}

\newcommand{\nineH}        {$\sqrt{s}=0.9$~Te\kern-.1emV}
\newcommand{\seven}        {$\sqrt{s}=7$~Te\kern-.1emV}
\newcommand{\twoH}         {$\sqrt{s}=0.2$~Te\kern-.1emV}
\newcommand{\twoHnn}         {$\sqrt{s_{\mathrm{NN}}}=0.2$~Te\kern-.1emV}
\newcommand{\twosevensix}  {$\sqrt{s}=2.76$~Te\kern-.1emV}
\newcommand{\five}         {$\sqrt{s}=5.02$~Te\kern-.1emV}
\newcommand{\thirdteennew}         {$\sqrt{s}=13.6$~Te\kern-.1emV}
\newcommand{\twosevensixnn}{$\sqrt{s_{\mathrm{NN}}}=2.76$~Te\kern-.1emV}
\newcommand{\fivenn}       {$\sqrt{s_{\mathrm{NN}}}=5.02$~Te\kern-.1emV}
\newcommand{\fivennnew}       {$\sqrt{s_{\mathrm{NN}}}=5.36$~Te\kern-.1emV}

\newcommand{\GeVc}         {Ge\kern-.1emV$/c$}
\newcommand{\MeVc}         {Me\kern-.1emV$/c$}
\newcommand{\TeV}          {Te\kern-.1emV}
\newcommand{\GeV}          {Ge\kern-.1emV}
\newcommand{\MeV}          {Me\kern-.1emV}
\newcommand{\GeVmass}      {Ge\kern-.1emV$/c^2$}
\newcommand{\MeVmass}      {Me\kern-.2emV$/c^2$}

\jname{Annual Review of Nuclear and Particle Science}
\jvol{75:463-486}
\jyear{2025}
\doi{10.1146/annurev-nucl-121423-100858}

\begin{document}

\markboth{R. Bailhache and H. Appelshäuser }{Dileptons at Colliders as Probes of the Quark-Gluon Plasma}

\title{Dileptons at Colliders as Probes of the Quark-Gluon Plasma}

\author{ R. Bailhache,$^1$  and H. Appelshäuser$^2$ 
\affil{$^1$Physik/Institut für Kernphysik, Goethe-Universität, Frankfurt a. Main, Germany, 60439; email: rbailhache@ikf.uni-frankfurt.de}
\affil{$^2$Physik/Institut für Kernphysik, Goethe-Universität, Frankfurt a. Main, Germany, 60439; email: appels@ikf.uni-frankfurt.de}}

\begin{abstract}
Ultra-relativistic heavy-ion collisions are used to create a deconfined state of quarks and gluons, the quark-gluon plasma (QGP), similar to the matter in the early universe. Dileptons are a unique probe of the QGP. Being emitted during all stages of the collision without interacting strongly with the surrounding matter, they carry undistorted information about the medium evolution. The mass of the lepton-antilepton pair gives a unique mean to separate partonic from hadronic radiation. Thus, dileptons can be used to study the QGP equilibration time, its average temperature but also effects related to the restoration of chiral symmetry in the hot medium via vector meson decays. This information is not accessible with hadrons. The price to pay is a large background from ordinary hadron decays. We summarize the potential of dilepton measurements, the results obtained so far at colliders, and the ongoing efforts for future experiments with further increased sensitivity.
\end{abstract}

\maketitle

\tableofcontents

\section{Introduction}

During the first few microseconds after the Big Bang, the universe is believed to be filled by an intriguing state of strongly interacting matter, the quark-gluon plasma (QGP). At high temperature ($T$) and small net-baryon density  ($\mu_{\rm B}$) the theory of the strong interaction, quantum chromodynamics (QCD), predicts a transition from ordinary matter made of hadrons to a plasma in which colored quarks and gluons are deconfined and determine the number of degrees of freedom\,\cite{Bazavov:2011nk,Borsanyi:2013bia,HotQCD:2018pds,Borsanyi:2020fev}. Ultra-relativistic heavy-ion collisions are the only known way to recreate this system of deconfined partons in the laboratory and to study its properties. By varying the center-of-mass energy per nucleon pair of the collision ($\sqrt{s_{\rm NN}}$), the phase diagram of strongly interacting matter can be explored from large net-baryon density ($\mu_{\rm B}$) and moderate temperature at $\sqrt{s_{\rm NN}}$ of some GeV to small net-baryon density and high temperature for $\sqrt{s_{\rm NN}}$ of a few TeV. Collisions of lead ions at the Large Hadron Collider (LHC) at CERN produce the hottest and longest-lived medium ever created in the laboratory with $\mu_{\rm B}\simeq 0$. The temperature is expected to exceed those in stellar interiors by about a factor of 10$^{5}$.\\  

Evidence for the formation of strongly interacting matter at unprecedented temperatures and energy densities was derived from the observation of collective flow in Au--Au collisions at the top energy ($\sqrt{s_{\rm NN}} = 200\,$GeV) of the Relativistic Heavy Ion Collider (RHIC). Measurements of the azimuthal momentum anisotropy of hadrons emerging from non-central heavy-ion collisions have revealed that the created plasma behaves like an almost perfect fluid, with extremely low viscosity to entropy density ratio\,\cite{Lacey:2001va,Shuryak:2004cy,Gyulassy:2004zy}. Contrary to the naive expectations of a weakly interacting gas of quarks and gluons, calculations assuming a strongly coupled QGP (sQGP) describe the data best. However, hadrons decouple at the end of the system evolution long after the sQGP reached thermal equilibration and information about the early stage of the collision is washed out by subsequent interactions. Therefore, a model-dependent extrapolation of system properties is necessary to extract some knowledge about the early partonic phase of the heavy-ion collision from final-state hadron measurements.

\section{Dileptons at colliders - opportunities and challenges}

Photons and dileptons, i.e. charged lepton-antilepton pairs arising from the internal conversion of virtual photons, have the unique property of being produced in all stages of heavy-ion collisions, especially during the hot partonic and hadronic phases, and of not interacting strongly with the surrounding matter. This means that they contain undistorted information about the medium at the time of their emission, without being affected by significant subsequent interactions during the system evolution. Compared to real photons, dileptons have an additional degree of freedom, the invariant mass $m_{\rm ll}$ of the lepton-antilepton pair ($\rm e^{+}e^{-}$ or $\rm \mu^{+}\mu^{-}$). Different mass ranges typically correspond to different production times. From high to low mass, dileptons are emitted on average in the early to late stages of the heavy-ion collision. In particular, $\rm l^{+}l^{-}$ pairs with invariant mass greater than about 1.2\,GeV/$c^{2}$ provide unique access to the radiation from the early partonic phase. They are well separated from the radiation from the hadronic phase, which is characterized by smaller dilepton invariant masses\,\cite{Rapp:2014hha}.

\subsection{Shining messenger of the early medium}

Since $\rm l^{+}l^{-}$ pairs are generated by different sources from the beginning to the end of the collision, dilepton measurements offer the unique possibility to shed light on the temporal evolution of the medium, including the thermalization mechanism, the time dependence of the temperature inside the fireball, and a possible partial restoration of chiral symmetry in the hot hadronic phase. \\

The QCD Lagrangian is symmetric under chiral transformation between left- and right-handed states for massless quarks\,\cite{Bilic:1997sh,Dominguez:2012bs}. In hadronic matter, chiral symmetry would imply that left- and right-handed quarks are interchangeable in bound states of mesons and baryons. Every state would appear in a pair of equal mass and opposite parity particles, called "chiral partners". The mass spectrum of hadrons, however, indicates that this is not the case and chiral symmetry is broken, not only explicitly by the quark masses but also spontaneously with the formation of a chiral condensate ($\langle {\rm q\bar{q}} \rangle\neq 0$). The QCD vacuum is in fact not an empty space but filled, among other things, with quark-antiquark pairs, which break chiral symmetry. Real quarks and antiquarks interact with these quark and gluon condensates and generate most of the mass of ordinary matter. In comparison, the contribution of the mass generated by the Higgs field is almost negligible. However, at high temperatures or high matter densities, where the expectation value of the chiral condensate is predicted to decrease, chiral symmetry is expected to be restored. Lattice QCD (lQCD) calculations show that chiral symmetry should be partially restored in very hot hadronic matter\,\cite{Borsanyi:2010bp,Bazavov:2011nk,Borsanyi:2020fev}, as it arises when the medium cools in the later stages of ultra-relativistic heavy-ion collisions. Chiral partners, i.e. pairs of hadrons with same spin and quark content but opposite parity like vector mesons ($\rm \rho, \omega$ and $\rm \phi$) and their axialvector counterparts ($\rm a_{1}(1260), f_{1}(1285)$ and $\rm f_{1}(1420)$), have then similar masses and mix. Whereas it is very challenging to measure the properties of axialvector mesons in heavy-ion collisions, vector mesons decay directly into dileptons, which can be easily reconstructed in the detectors. The $\rm \rho$ is by far the most interesting one because it strongly couples to the $\pi^{+}\pi^{-}$ channel and has a lifetime of only 1.3\,fm/$c$, much shorter than that of the hot fireball produced at RHIC or at the LHC. As a consequence, $\rm{\rho}$ mesons are expected to be copiously produced in the vicinity of the chiral phase transition with a modified spectral function broadened by scattering off baryons in the dense hadronic matter\,\cite{Pisarski:1995xu,Rapp:1999us,vanHees:2007th,Hohler:2013eba}. Although a direct measurement of both chiral partners would allow an unambiguous signature of chiral symmetry restoration, first-principle Weinberg sum rules which relate moments of the difference between the vector and axialvector spectral functions to chiral order parameters provide a direct link between the $\rho$ meson spectral function and the restoration of chiral symmetry\,\cite{Hohler:2013eba}. Therefore, dilepton invariant mass spectra at masses below 1.2\,GeV/$c^{2}$ allow the measurement of spectral functions of vector mesons in the medium, in particular of the $\rm \rho$, and the study of the mechanisms for restoring chiral symmetry in the hot medium in which they are produced.\\

The production of $\rho$ mesons in the hot fireball can be interpreted as thermal radiation at the phase boundary and from the hadronic phase. At larger mass, between about 1 and 3\,\GeVc, thermal radiation is expected to originate predominantly from the QGP and to be sensitive to the equation of state in the QGP. An estimate of the initial QGP temperature can be derived from its exponential invariant mass distribution. The correlation between the slope parameter of the thermal spectrum and the initial temperature was first shown within the expanding fireball model of R. Rapp {\it{et al.}}\,\cite{Rapp:2014hha} using a lQCD inspired equation of state for the QGP. The measured temperature is likely to be somewhat lower than the actual temperature but has the unique property of being neither distorted by blueshift effects due to the collective expansion of the medium, such as the slope parameters obtained from transverse
momentum (\pt) spectra, nor limited by the hadron formation temperature, as in temperatures determined using hadronic final-state \pt ~distributions or yields. The potential of QGP thermal radiation measurements as fireball thermometer was confirmed recently by computations of thermal QCD dilepton production rates at next-to-leading order and finite net-baryon density integrated over space-time using viscous hydrodynamics\,\cite{Churchill:2023zkk}. In these calculations, the QGP thermal yield in the intermediate mass range (1-3\,\GeVmass) is found to offer a quantitative assessment of the lifetime of the fireball\,\cite{Churchill:2023vpt}, similarly to the dilepton thermal yield at lower mass (0.3-0.7\,\GeVmass) initially suggested in Ref.\,\cite{Rapp:2014hha}. However, radiation from the partonic medium before the plasma reaches local thermal equilibrium may play a role already at masses of the order of 2-3\,\GeVmass.\\

Thermalization processes have not yet been well studied, although there is experimental evidence, especially from hadronic final-state observables, that the medium produced in heavy-ion collisions reaches sufficient equilibrium to be described by relativistic viscous hydrodynamics. Dileptons emitted from the pre-equilibrium phase are a unique tool for studying the onset of thermalization\,\cite{Coquet:2021lca,Coquet:2021gms,Coquet:2023wjk}. They are expected to be best experimentally accessible in a \mll\,\,window around 2-4\,\GeVmass. At lower and higher mass, they are expected to be dominated by thermal QGP dileptons and Drell-Yan pairs, respectively. Since the polarization of the $\rm l^{+}l^{-}$ pairs depends on their production process, measurements of the angular distribution of dileptons are instrumental in identifying this invariant mass window and providing information on the equilibration time of the QGP, which is directly related to the shear viscosity to entropy ratio $\eta/s$ in the early stages of the collision\,\cite{Coquet:2023wjk}.\\

Thanks to the diversity of dilepton sources and the relationship between their invariant mass and emission times, multidifferential analyses of the $\rm l^{+}l^{-}$ production open up the possibility to investigate the time evolution of the medium properties in the collisions, such as the radial flow with the pair transverse momentum \ptll, the elliptic flow (\vtwoll)\,\cite{Vujanovic:2019yih,Kasmaei:2018oag}, and the temperature of the system\,\cite{Rapp:2014hha}. This makes them important probes for the transport coefficients in the earliest phases of the collision, not only for $\eta/s$\,\cite{Coquet:2023wjk} but also the bulk viscosity to entropy ratio $\zeta/s$ via \vtwoll\,\cite{Vujanovic:2019yih} or the electric conductivity $\sigma_{\rm EM}$. The latter determines the diffusion of electric charges in the medium and their response to electric fields. Although detailed  knowledge of $\sigma_{\rm EM}$ is required to understand phenomena related to the presence of strong magnetic fields in the early stages of heavy-ion collisions\,\cite{Huang:2015oca,Tuchin:2015oka}, such as the Chiral Magnetic Effect, theoretical calculations show a large spread of the expected $\sigma_{\rm EM}$ temperature dependence without any experimental constraints\,\cite{Fotakis:2021diq}. In the limit of small energies (\mll $\simeq$ \ptll $\simeq$ 0), the predicted yield of thermal dileptons is sensitive to $\sigma_{\rm EM}$\,\cite{Ding:2016hua}. Different electric conductivities lead to different amplitude and width of the so-called conductivity peak of the spectral function\,\cite{Moore:2006qn,Floerchinger:2021xhb,Rapp:2024grb}. Therefore, measurements of thermal radiation at very low masses and momenta can provide
experimental constraints on $\sigma_{\rm EM}$. Note that in the zero mass limit, the fraction of direct to inclusive virtual photons is also equivalent to the same fraction for real photons. Consequently, the measured dilepton yield in the quasi-real virtual-photon region, where \ptll\,\, is much larger than \mll, can be used to measure the \pt-differential yield of direct photons, i.e. photons not originating from hadronic decays\,\cite{PHENIX:2009gyd}.\\ 

Finally, at very low pair transverse momentum, another electromagnetic process contributes significantly to dilepton production in peripheral and ultra-peripheral heavy-ion collisions. Quasi-real photons, arising from the highly Lorentz-contracted electromagnetic fields of the two incoming nuclei can interact via the Breit–Wheeler process\,\cite{Breit:1934zz} and produce $\rm l^{+}l^{-}$ pairs with characteristically small \ptll. Such a production mechanism is not tied to the occurrence of a hadronic heavy-ion collision with the formation of a QGP. Nevertheless, the observation of photon–photon production of dileptons in mid-central and central heavy-ion collisions at RHIC\,\cite{STAR:2018ldd} and at the LHC\,\cite{ATLAS:2018pfw,ATLAS:2022yad,ALICE:2022hvk} raised the question if they could not be affected by the presence of the hot medium and thus serve as a messenger of the early system properties. Since recent theoretical developments significantly limit the scope for any such medium-induced effects\,\cite{Brandenburg:2021lnj}, this process will not be part of this review.


\subsection{State-of-the-art models}
\label{models}

Results on dilepton production from fixed-target experiments at the CERN-SPS (NA38\,\cite{NA38:1993pef}, HELIOS-3\,\cite{HELIOS3:a, HELIOS3:1998xeb}, CERES\,\cite{CERES:1995vll, CERESNA45:1997tgc, CERESNA45:2002gnc, CERES:2006wcq}, NA60\,\cite{NA60:2006ymb, NA60:2008dcb, NA60:2008ctj, Specht:2010xu}) triggered intensive theoretical discussions on different dilepton production scenarios. Two main approaches to estimate the modification of the spectral functions of vector mesons (especially the short-lived $\rho$) produced in the very hot hadronic medium, where chiral symmetry is expected to be partially restored, were established at that time. One predicted a drop of the pole mass of vector (and axialvector) mesons due to a $T$- and $\mu_{\rm B}$-dependent reduction of the quark condensate $\langle \rm q\bar{q}\rangle$\,\cite{Brown:1995qt}. The other resulted in an effective broadening of the spectral function owing to many-body collisions in the vector-meson dominance picture\,\cite{Rapp:1999us, Rapp:1999ej, Eletsky:2001bb}. Both scenarios led to an increased contribution of $\rm l^{+}l^{-}$ pairs from thermally produced $\rho$ mesons with an enhanced dilepton yield below the $\rho$ mass compared to measurements in vacuum, i.e. proton-induced collisions. In fact, the SPS experiments reported significant excess radiation of dileptons below the $\rho$ mass region in ultra-relativistic heavy-ion collisions at around $\snn\approx 20$\,GeV, beyond final-state hadron decays. Precision measurements from the NA60\,\cite{NA60:2006ymb} and CERES\,\cite{CERES:2006wcq} experiments demonstrated that the excess is consistent with thermal radiation from a locally equilibrated fireball, with the low-mass spectra requiring a broadening of the $\rho$ spectral function, while the dropping mass scenario failed to describe the data. For $\mll > 1$\,GeV, the observed large sudden drop of the inverse slope of the dilepton \pt\,\,as a function of invariant mass indicates that the excess $\rm l^{+}l^{-}$ pairs originate from the partonic phase, where radial flow did not build up yet\,\cite{NA60:2008ctj}. A transition to dominantly partonic emission sources in this mass region was supported by the observation of an average temperature of about 200\,MeV associated with the excess mass spectrum over the range $1.2 < \mll < 2$\,\GeVmass\,\cite{Specht:2010xu}. Together, these measurements gave important constraints to the models which, at that time, predicted different relative contributions for radiation from the partonic phase at $\mll > 1$\,\GeVmass.\\

More recently, there are two main categories of models for thermal dilepton production in heavy-ion collisions: models based on effective many-body theory and microscopic dynamic transport models. 

In the effective many-body theory models, the thermal dilepton production rate is computed from the imaginary part of the retarded photon self-energy, i.e. the electromagnetic spectral function, assuming local thermal equilibrium of the fluid elements of an expanding fireball. In hadronic matter, based on the vector-meson dominance model (VDM), the retarded photon self-energy turns into a vector–meson propagator in the medium. The $\rho$-meson propagator is computed from interactions of the $\rho$ with mesons and baryons. The resulting broadening of the $\rho$ spectral function is mostly attributable to the interactions with baryons rather than mesons\,\cite{Rapp:1999ej, Rapp:2009yu, Xu:2011tz, Vujanovic:2013jpa}. Therefore, the thermal dilepton yield in the hadronic phase of heavy-ion collision is very sensitive to the total baryon and antibaryon density (rather than the net-baryon density) in the medium. In the QGP, the thermal radiation rates are calculated using constraints from lattice QCD calculations above the critical deconfinement temperature $T_{\rm c}$\,\cite{Rapp:2014hha, Ding:2016hua}. Alternatively, it can be computed at up to next-to-leading order perturbatively in the strong coupling, as done recently for finite baryon chemical potential\,\cite{Churchill:2023zkk, Churchill:2023vpt}. The QGP and in-medium hadronic rates are nearly degenerate at temperature around $T_{\rm c}$. This is referred to as the "parton-hadron" duality\,\cite{Rapp:2009yu}. The final thermal dilepton yields are obtained from the integrated radiation rates over the full space-time evolution of the medium. Models differ in the way they implement the space-time evolution. In the approach of Rapp\,\cite{vanHees:2007th, Rapp:2013nxa}, a cylindrical expanding fireball is used with a lattice QCD equation of state above $T_{\rm c}$ and a hadron resonance gas below. Alternatively, the space-time evolution can also be obtained from viscous hydrodynamical simulations tuned to reproduce experimental hadron data\,\cite{Churchill:2023zkk, Vujanovic:2019yih, Kasmaei:2018oag}. The sensitivity of the dilepton yield and elliptic flow to the bulk\,\cite{Vujanovic:2019yih} or shear\,\cite{Kasmaei:2018oag} viscosity to entropy density ratios can be studied in this way.



In transport models, the different dilepton production channels in a matter that is a priori not necessary in equilibrium are investigated microscopically. One of such models is the parton-hadron string dynamic (PHSD) covariant transport model\,\cite{Cassing:2009vt, Bratkovskaya:2011wp, Song:2018xca}. In the partonic phase, $\rm l^{+}l^{-}$ pairs ($\rm \gamma^{*}$) are produced by the constituents of the strongly interacting QGP via $\rm q + \bar{q}$ Born annihilation, quark-gluon Compton scattering ($\rm q(\bar{q}) + g \to \gamma^{*} + q(\bar{q})$) and quark-antiquark annihilation with gluon Bremsstrahlung in the final state ($\rm q + \bar{q} \to g + \gamma^{*}$). The partonic channels are calculated assuming off-mass shell partons with a phenomenological parametrisation of the quark and gluon propagators in the QGP based on the dynamical quasiparticle model (DQPM) matched to reproduce lattice QCD results in thermodynamic equilibrium. In the hadronic phase, the dielectron production is computed using in-medium modified electromagnetic spectral functions of low-mass vector mesons which evolve dynamically towards on-shell spectral functions in the vacuum. 

Both approachs, effective many-body theory models and transport models, deliver similar predictions for the dilepton radiation yield in the mass region $0.4 < \mll < 0.9$\,\GeVmass, although the relative contribution from the partonic phase may be different.\\

Recently, the production of dileptons with high invariant mass ($\mll \geq 2.$\,\GeVmass) from the early non-equilibrium QGP has raised the interest of the community as a unique probe of this very early phase of the heavy-ion collision, which stays inaccessible with other observables. Before about $1$\,fm/$c$, where the system reaches sufficient equilibrium to be described by relativistic viscous hydrodynamics, the partonic medium is gluon dominated and highly anisotropic due to the rapid longitudinal expansion. QCD kinetic theory provides a robust way to model the pre-equilibrium dynamics\,\cite{Giacalone:2019ldn, Kurkela:2018wud, Kurkela:2018xxd, Du:2020zqg}. The information about the thermalization of the medium is contained in a single parameter, typically the early shear viscosity to entropy density ratio. Such theory has been used to study the sensitivity of the pre-equilibrium dilepton production to the kinetic and chemical equilibration time of the quark-gluon plasma\,\cite{Coquet:2021lca, Coquet:2021gms, Coquet:2023wjk, Garcia-Montero:2024msw}.


\subsection{Experimental challenges}

Dileptons have a unique potential to probe the early medium formed in ultra-relativistic heavy-ion collisions. However, the $\rm l^{+}l^{-}$ pairs emitted from the partonic and hot hadronic phases are difficult to measure.\\

First, the production cross sections for thermal or pre-equilibrium radiation from the medium are small. The internal conversion probability of the virtual photon ($\sim10^{-2}$) together with the rapidly decreasing cross section as a function of \mll\,\,($\propto  1/\mll$) make early radiation from the hot fireball a rare probe. Compared to hadrons, leptons are produced more than two orders of magnitude less frequently. Moreover, most of these leptons are originating from hadronic decays happening long after the medium cooled down and decoupled, only a small fraction is part of the dilepton signal. Therefore, dilepton measurements require detectors with high quality particle identification capabilities, large acceptance and high readout rates to allow for the inspection of a large number of events.\\

Second, experiments necessarily integrate the $\rm l^{+}l^{-}$ pairs emitted throughout the full evolution of the heavy-ion collision where the dilepton signal of interest, i.e.\,radiation from the partonic and hot hadronic phases, represents only a tiny component. A large physics background of $\rm l^{+}l^{-}$ pairs produced much later in the collision via hadron decays must be taken into account. In particular, dileptons from correlated open heavy-flavour hadron decays overwhelm the QGP and pre-equilibrium signal in the intermediate mass range $1.2 < \mll < 5$\,\GeVmass~at the top RHIC and LHC energies ($\snn \geq 0.2$\,TeV)\,\cite{STAR:2015tnn,PHENIX:2015vek,ALICE:2018ael,ALICE:2023jef}. They are produced in semi-leptonic decays of open heavy-flavour hadrons originating from a single hard-scattering process, i.e.\,the same quark-antiquark pair ($\rm c\bar{c}$ or $\rm b\bar{b}$), and are correlated via flavour conservation. Whereas such background is small at SPS energies ($\snn$ of a few GeV)\,\cite{NA60:2008dcb}, it increases significantly towards top RHIC and LHC energies due to the steep energy dependence of the heavy-flavour production cross sections. 

Measurements of individual hadronic final states are usually used to estimate the contributions of hadronic decays. Parameterisations of the $p_{\rm T}$-differential yields of the different mother hadrons are taken as input for a fast Monte Carlo simulation that performs the decays and allows the determination of the corresponding expected dilepton yields\,\cite{STAR:2015tnn,PHENIX:2015vek}. The hadrons may never be directly measured, in which case an extrapolation from other hadron measurements using $m_{\rm T}$-scaling or Blast-wave fits is necessary. However, such a technique is difficult to apply to the contribution of correlated open hadron decays with heavy flavour. Since electron and positron do not originate from the same hadron, additional knowledge of the correlation between the two open heavy-flavour hadrons is required. The latter as well as the $p_{\rm T}$ spectrum of the individual hadrons are influenced by the presence of the hot fireball due to strong interactions of the heavy quarks with the plasma and possible further scattering of the generated open heavy-flavour hadrons in the hot hadronic medium. Therefore, additional experimental means are required to separate the heavy-flavour background from the early radiation in ultra-relativistic heavy-ion collisions. 

At higher masses, $\mll > 3-4$\,\GeVmass\,\,at the LHC or lower for smaller $\sqrt{s_{\rm NN}}$\,\cite{Coquet:2021lca}, dileptons produced via the Drell-Yan process are expected to dominate the partonic radiation from the early phase of the collision. The calculated Drell-Yan production cross section suffers from large theoretical uncertainties in the low-mass
regime due to its non-perturbative nature. The polarization of the $\rm l^{+}l^{-}$ pairs could nevertheless be used as a discriminating variable to disentangle the different sources of dileptons\,\cite{Coquet:2023wjk}. 

An additional physics background of $\rm e^{+}e^{-}$ pairs produced via photon conversion in the detector material is affecting the dielectron measurements at very low masses. The reconstructed invariant mass of such pairs at the collision point is typically biased towards non-zero finite values, due to the extrapolation of the tracks in the magnetic field used in the experiment from the detector layer where the conversion happened to the primary vertex. Depending on the position of the specific detector layer, i.e.\,its radial distance from the beam axis, such conversion $\rm e^{+}e^{-}$ pairs can contaminate the dilepton spectrum typically up to $\mee\leq$ 0.2\,\GeVmass. However, they can be easily identified thanks to the characteristic orientation of their opening angle relative to the magnetic field\,\cite{PHENIX:2009gyd}. Nevertheless, it is imperative to suppress their contribution by minimizing the material near the beam axis, since they contribute not only to the physical background but also to the combinatorial background.

Finally, since it is not possible to identify which lepton belongs to which antilepton, a statistical approach is used to extract the dilepton signal ($S$). All identified leptons are combined to form the distribution of same-event pairs of opposite sign. The latter consists of true signal pairs as well as background pairs ($B$). The background pairs are mainly combinatorial but also contain residual correlations such as those arising from jets and from conversions of correlated decay photons originating from the same mother particle. The background yield roughly scales with the square of the charged-particle multiplicity in the detector acceptance window and becomes therefore very large at high \snn ~and in central heavy-ion collisions. The signal-to-background ratio ($S/B$) can be as small as $10^{-3}$ in central Au--Au collisions at \twoHnn\,\cite{STAR:2015tnn,PHENIX:2015vek} or even down to $10^{-4}$ in central Pb--Pb collisions at \five\,\cite{ALICE:2018ael,ALICE:2023jef}. For this reason, the combinatorial background must be  estimated with high precision and at the same time reduced as much as possible. Different methods are used. The same-sign pairs from the same events provide the best self-normalised estimate of the combinatorial and correlated backgrounds, as long as they are charge symmetric, i.e.\,they produce the same yield and distribution of same- and opposite-sign pairs. While this approach, the so-called like-sign method, works well for detectors with uniform $\rm 2\pi$ azimuthal coverage\,\cite{STAR:2015tnn}, these conditions are generally not met in spectrometers with limited azimuthal acceptance due to substantial acceptance difference between same- and opposite-sign pairs\,\cite{PHENIX:2015vek}. Furthermore, poor lepton purity may also let the procedure break down, since lepton-misidentified hadron pairs are then susceptible to have different distributions depending on their total charge. Finally, at high \mll, the like-sign method may suffer from large statistical uncertainties.  An alternative way to estimate the background is to calculate the different correlated background contributions and evaluate separately the combinatorial background with opposite-sign pairs from different events, having similar bulk properties (orientation of the reaction plane, charged-particle multiplicity $\cdots$)\,\cite{PHENIX:2015vek}. It is then crucial to have not only the pure detector effects under control, but also an accurate description of the underlying physics, such as jets, hadron spectra and correlations.


\section{Results from RHIC} 

\subsection{Top RHIC energy: 0.2\,TeV}

The first results on dilepton measurements at RHIC were reported by the PHENIX Collaboration with the production of $\rm e^{+}e^{-}$ pairs at midrapidity in Au--Au collisions at \twoHnn\,\cite{PHENIX:2009gyd}. Au--Au and pp collisions recorded at the same center-of-mass energy in 2004 and 2005, respectively, were analysed. On the one hand, the dielectron yield in pp collisions could be explained by known contributions from light meson decays, heavy-flavour hadron decays, and virtual direct photons, the latter being consistent with NLO pQCD calculations. On the other hand, a very large excess compared to the expectation from the hadronic cocktail was observed in Au--Au collisions at low mass $0.15 < \mee < 0.75$\,\GeVmass, in particular at low pair transverse momentum \ptee\,\,($\ptee < 1$\,\GeVc). All models that successfully reproduced dilepton measurements at smaller \snn\,\, from the SPS fixed target experiments\,\cite{PHENIX:2009gyd,Linnyk:2011vx} failed to explain the PHENIX data. In 2014, the STAR Collaboration published their first dielectron spectra in Au--Au collisions at midrapidity and at the same \snn\,\,based on data recorded in 2010 and could not confirm the finding of PHENIX\,\cite{STAR:2013pwb}. The enhancement reported by STAR was much smaller, by a factor of about 2-3.

Whereas the PHENIX detector consists of two spectrometers with limited azimuthal acceptance, STAR benefits from a time projection chamber (TPC) and a time-of-flight (TOF) detector with a 2$\pi$ azimuthal coverage. This greatly simplifies the estimation of the combinatorial and correlated background. First, the combination of both detectors enables a very high electron purity, which is above 92\% in central Au--Au collisions\,\cite{STAR:2013pwb, STAR:2015tnn} compared to 70\% for the first PHENIX paper\,\cite{PHENIX:2009gyd} relying on ring-imaging cherenkov (RICH) detectors and calorimeters. Second, the acceptance is similar for same- and opposite-sign lepton pairs, and therefore the like-sign method can be safely utilized to determine the sum of the combinatorial and correlated background with only small corrections. The inconsistency in the measurements of the low-mass excess between the two experiments was solved with the new data from PHENIX, collected in 2010 after the installation of the Hadron Blind Detector (HBD)\,\cite{PHENIX:2015vek}. The latter helps to improve significantly the electron purity using a neuronal network approach and the signals in the calorimeters, HBD, TOF and RICH detectors. An extensive study of the combinatorial and correlated background shows that the contribution of misidentified hadrons paired with electrons leads to very different same-sign and opposite-sign distributions, breaking the validity of the like-sign method. Moreover, due to the limited acceptance of the two PHENIX spectrometers in azimuth, same- and opposite-sign electron pairs have very different acceptance, giving rise to large correction factors. Such corrections are sensitive for example to imperfect description of effects related to the collective flow in the heavy-ion collisions because of the finite resolution of the reconstructed event plane with the PHENIX detector\,\cite{PHENIX:2015vek}. Any relative uncertainty on the combinatorial and correlated background propagates then to the final dilepton spectrum with a factor $B/S$. At RHIC, it can be multiplied by a factor of up to 10$^{3}$ depending on the centrality of the collision, and mass and \pt\,\,of the $\rm l^{+}l^{-}$ pairs. Very unfortunately, the signal-to-background ratio is the smallest around masses of about 0.5\,\GeVmass, exactly where a significant contribution of thermally produced $\rho$ mesons in the hot hadronic medium is expected and where the very large excess was reported in the first PHENIX measurement. This demonstrates the importance of high lepton identification capability, large azimuthal acceptance, and a reconstruction algorithm that is robust to track overlap in an environment of high charged-particle multiplicity.\\

\begin{figure}[h]
\begin{minipage}{0.5\textwidth}
  \centering
  \includegraphics[width=1.\linewidth]{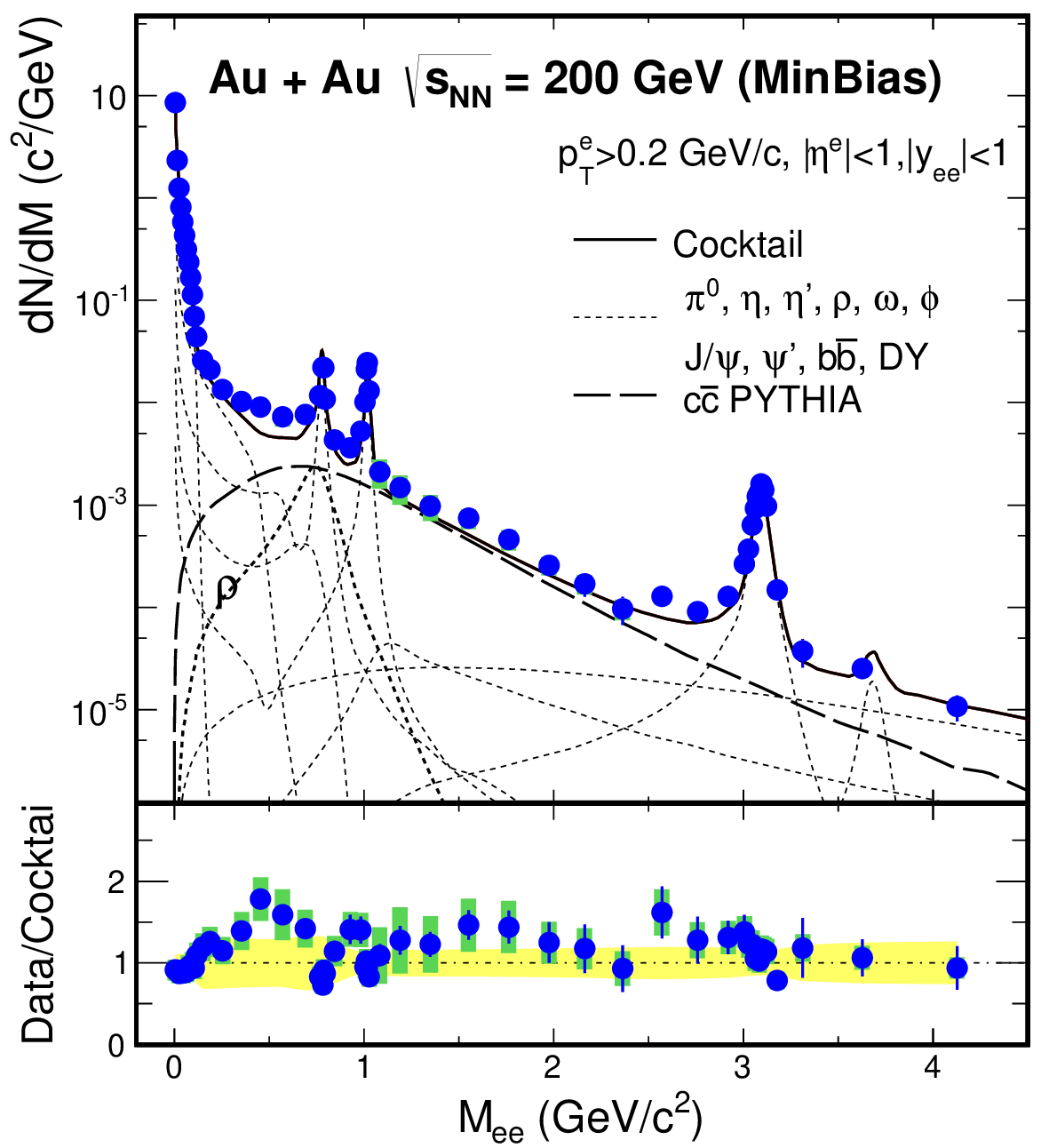}
\end{minipage}%
\begin{minipage}{.5\textwidth}
  \centering
  \includegraphics[width=1.\linewidth]{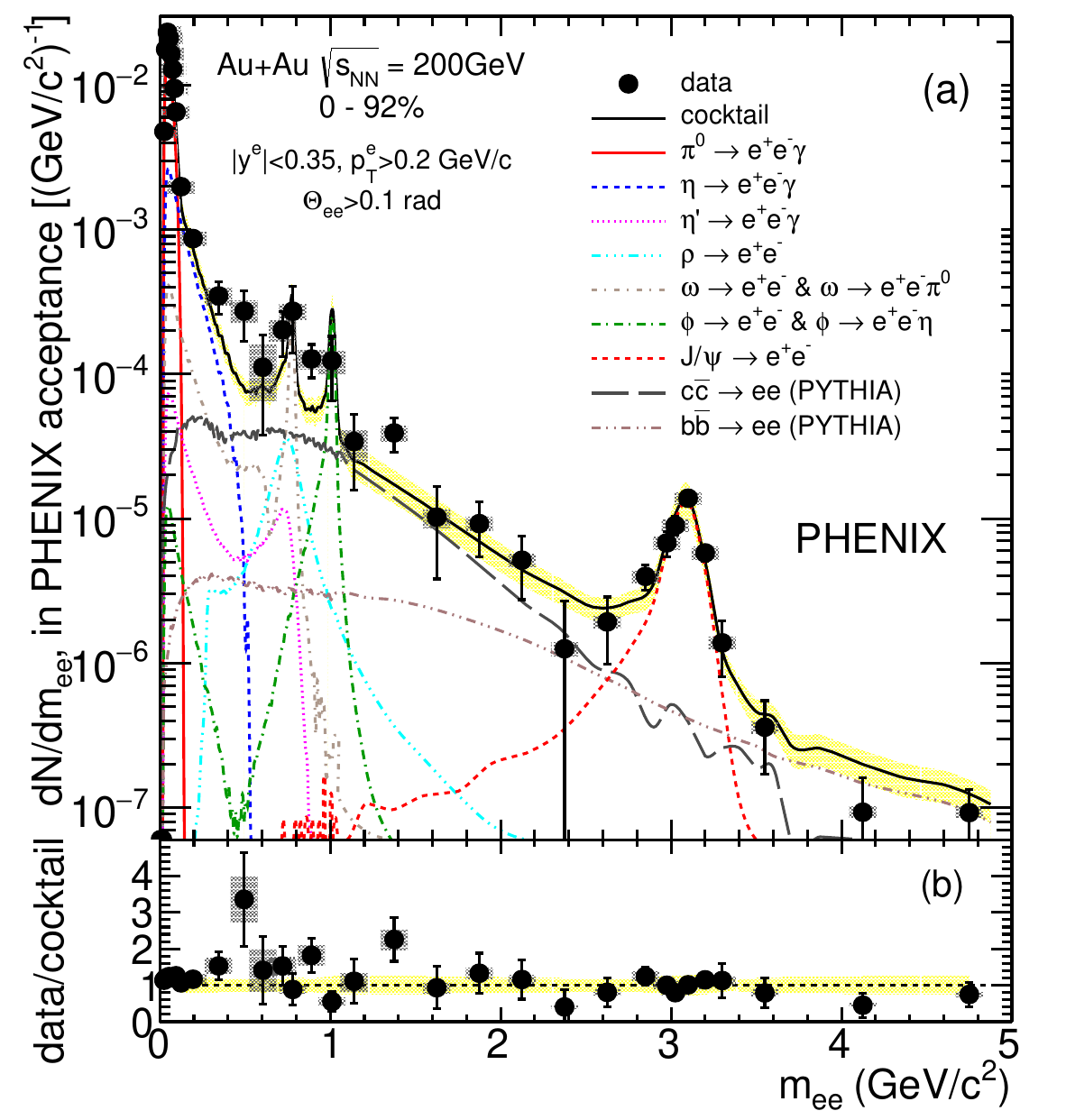}
\end{minipage}
\caption{Measured dielectron yields a a function of invariant mass in the STAR (left panel from Reference\,\cite{STAR:2015tnn}) and PHENIX (right panel from Reference\,\cite{PHENIX:2015vek}) acceptances in minimum bias Au--Au collisions at \twoHnn, compared to the contributions of known hadronic decays (cocktail). Statistical and systematic uncertainties on the data points are shown separately by vertical bars and boxes, respectively. The ratios data over cocktail simulations are shown in the bottom panels, where the band around unity indicates the uncertainties on the cocktail calculations.\label{firstfigure}}
\end{figure}


In \textbf{Figure \ref{firstfigure}} the most recent dielectron measurements in the STAR (left)\,\cite{STAR:2015tnn} and PHENIX (right)\,\cite{PHENIX:2015vek} acceptances in minimum bias (0--80\% and 0--92\%, respectively) Au--Au collisions at \twoHnn\,\, are shown as a function of \mee. The spectra are subject to the same minimum \pt\, requirement on single electrons with a larger pseudorapidity acceptance for STAR compared to PHENIX, both being at midrapidity. In the case of PHENIX, the azimuthal angle of the single electrons is also limited to the ideal acceptance of the two spectrometer arms (not specified in the figure), and an additional minimum pair opening angle of 100\,mrad is applied. The data are compared to cocktails of expected hadronic decays, where the PYTHIA generator is used in both cases to calculate the correlated pairs from open heavy-flavour hadron decays. The simulations are normalised to different charm cross sections in pp collisions 
($\rm {d}\sigma_{\rm c\bar{c}}/{\rm{d}y}= 171 \pm 26 \,\mu $b (STAR) and $106 \pm 34\,\mu $b (PHENIX))
and scaled with the average number of binary nucleon-nucleon collisions in Au--Au collisions. 

At low mass, below 1\,\GeVmass, the $\pi^{0}$-Dalitz as well as $\omega$ and $\phi$ resonance peaks are well visible. Whereas the cocktails can describe the data in the peak regions, an enhancement of $\rm e^{+}e^{-}$ pairs with respect to the expected hadronic decay contributions is observed for $0.3 < \mee < 0.76$\,\GeVmass\,\,by both collaborations. In this mass region, the 
cocktail yields are saturated by the decays of light mesons ($\eta$, $\rho$, and $\omega$) and open-charm hadrons ($\rm c \bar{c}$). In both cocktails, no medium effect on the $\rho$ mesons has been considered, although a significant contribution of $\rm e^{+}e^{-}$ pairs originating from $\rho$ mesons produced thermally in the medium is expected. The observed enhancement factor depends on the generator used to evaluate the heavy-flavour yield, i.e. PYTHIA or MC@NLO, as shown by PHENIX\,\cite{PHENIX:2015vek}. Dielectron measurements in pp or d--Au collisions are used to constrain the contribution from correlated heavy-flavour hadron decays neglecting any medium effect\,\cite{PHENIX:2014edx, PHENIX:2017ztp}. However, at RHIC energies, the inclusive $\rm e^{+}e^{-}$ yield in such colliding systems is dominated by vector meson decays for \mee\,\, smaller than about 1\,\GeVmass\,\, and the contribution of heavy-flavour hadron decays can be clearly determined only for higher masses. Therefore, the shape of the heavy-flavour distribution in this specific mass region ($0.3 < \mee < 0.76$\,\GeVmass) may not be so well known even in small collision systems without any further experimental tools.

At higher masses, $1.1 < \mee < 2.9$\,\GeVmass, the calculated dielectron yields from hadronic decays are dominated by correlated pairs from semileptonic decays of open-charm hadrons, with a small contribution from open-beauty hadrons and an even smaller contribution from Drell-Yan, not considered by the PHENIX Collaboration. The STAR data tend to be at the upper edge of the calculation uncertainties for $\mee \leq 2$\,\GeVmass\,\,and suggest slightly steeper falling spectra particularly for more central Au--Au collisions (Fig.34 of Ref.\,\cite{STAR:2015tnn}). A similar observation was made by PHENIX in mid-central Au--Au collisions (see Fig. 29 of Ref.\cite{PHENIX:2015vek}). 
The simulations presented here assume no modification of the heavy-flavour contribution due to the presence of the medium or cold nuclear matter. However, high-\pt\,\,single electrons from heavy-flavour hadron decays and open-charm hadrons are clearly suppressed compared to pp collisions at RHIC\,\cite{PHENIX:2006iih, STAR:2014wif}. Lacking an unambiguous way to identify $\rm e^{+}e^{-}$ pairs from open heavy-flavour hadron decays, it is not possible to distinguish between different scenarios including the modification of the correlated charm contribution and/or the existence of other contributing sources like radiation from the partonic medium.\\

\begin{figure}[h]
\begin{minipage}{.5\textwidth}
  \centering
  \includegraphics[width=1.\linewidth]{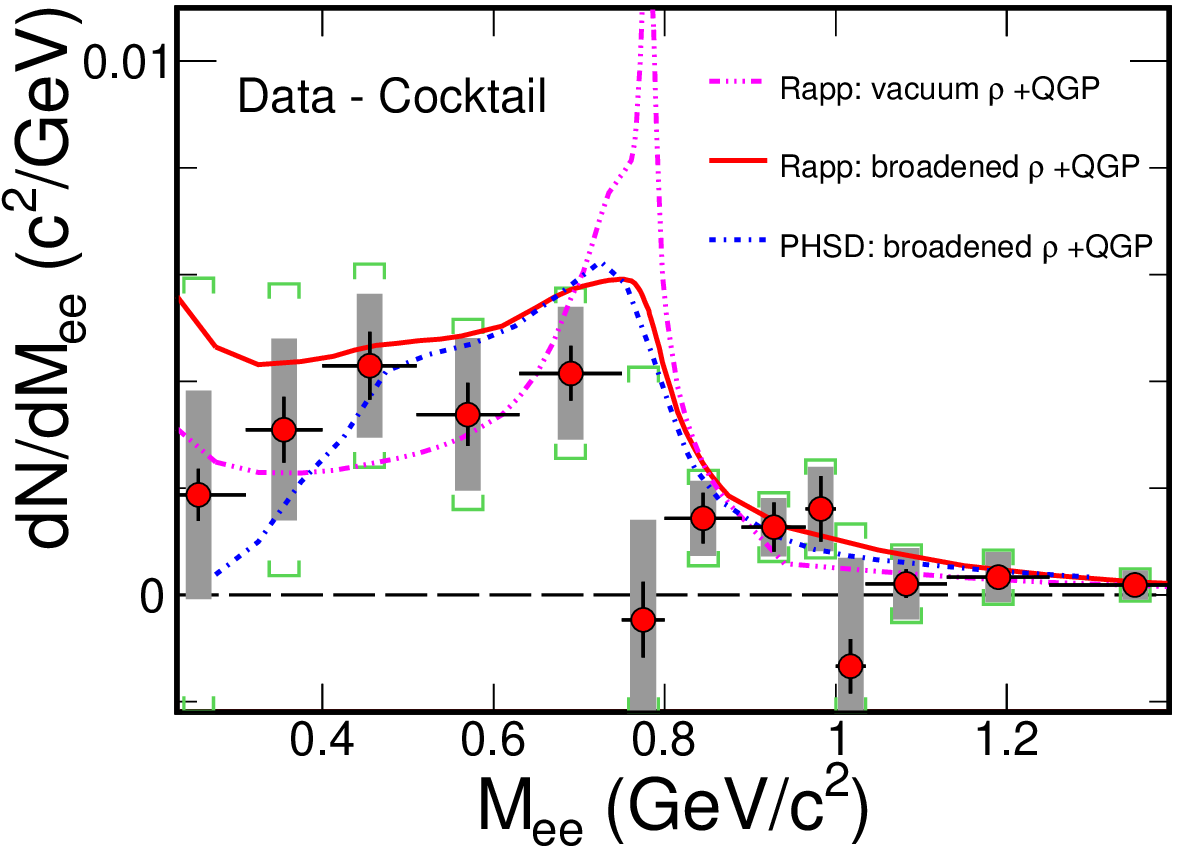}
\end{minipage}%
\begin{minipage}{.5\textwidth}
  \centering
  \includegraphics[width=1.\linewidth]{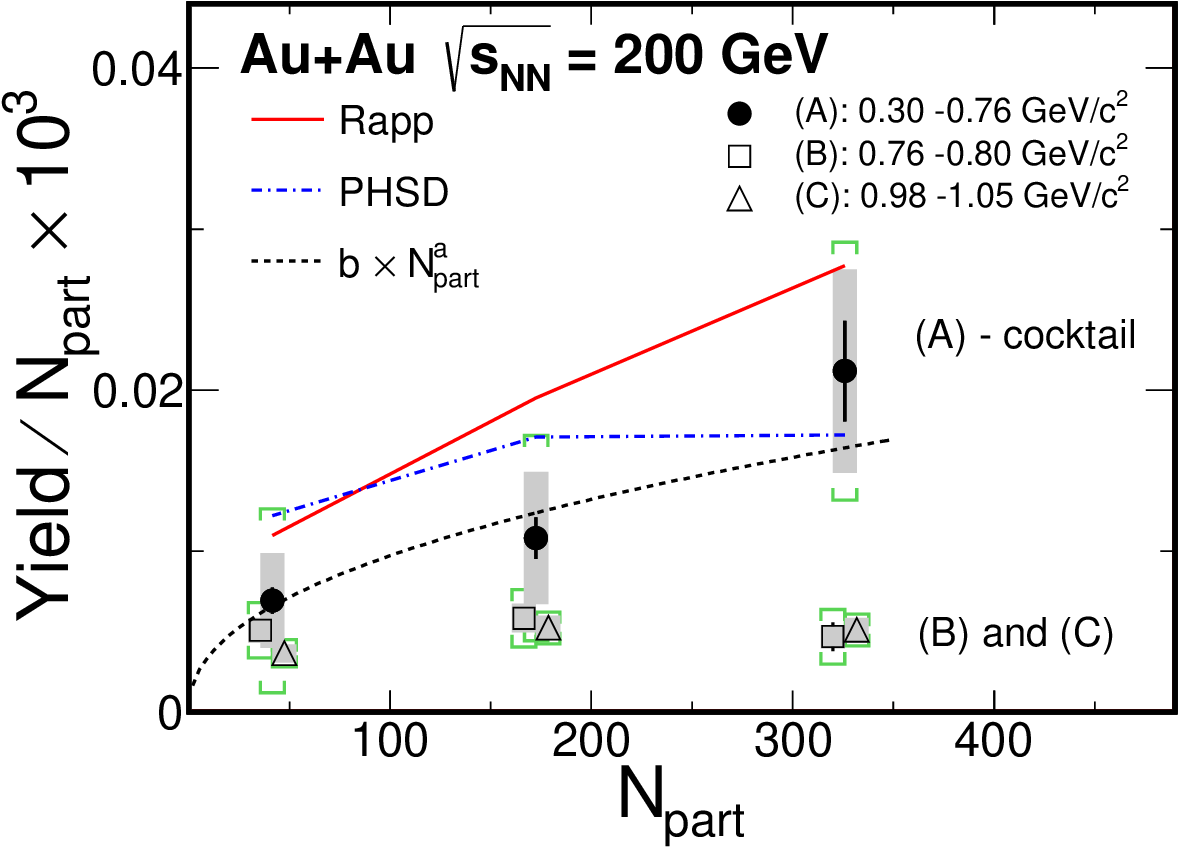}
\end{minipage}
\caption{Left panel: Invariant mass distribution of the dielectron excess (data - cocktail) in the STAR acceptance in minimum bias Au--Au collisions at \twoHnn\,\,compared to models. Panel from Reference\,\cite{STAR:2015tnn}. Right panel: Dielectron yields scaled by $N_{\rm part}$ for the $\rho$-like (A) region with the cocktail subtracted and for the $\omega$-like (B) and the $\phi$-like (C) regions without cocktail subtraction as a function of $N_{\rm part}$, together with model calculations (see text) and a fit of the data in the $\rho$-like region. Statistical and systematic uncertainties from the data points are shown separately by vertical bars and grey boxes, respectively. Green brackets depict the total systematic uncertainties including those from cocktails (to be ignored for B and C). Panel from Reference\,\cite{STAR:2015tnn}.
\label{secondfigure}}
\end{figure}

In the left panel of \textbf{Figure \ref{secondfigure}} the expected dielectron yield from known hadronic sources is subtracted from the STAR data to obtain the excess mass spectrum at low mass in minimum bias Au--Au collisions at \twoHnn\,\cite{STAR:2015tnn}. The $\rho$ contribution is not considered in the cocktail in order to guarantee a consistent comparison with calculations for thermal radiation from the hadronic and partonic phases, including $\rho$ produced during the full evolution of the system. Both the data and the computations are within the STAR acceptance. Two predictions are from the effective many-body model of Rapp\,\cite{Rapp:2013nxa} (see subsection \ref{models}) with a cylindrical expanding fireball. Above $T_{\rm c} = 170\,$MeV, the dielectron production is calculated using a lattice-QCD inspired approach. Below $T_{\rm c}$, the thermal emission rate of dielectrons from the hadronic phase is estimated via electromagnetic correlators based on the vector-meson dominance model approach with a chemical freeze-out temperature of $T_{\rm ch} = 160\,$MeV. The charged-particle multiplicity at midrapidity is tuned to reproduce the data. The two predictions differ by the choice of the $\rho$ spectral function. In one case, it is estimated in vacuum from an effective $\pi\rho$ Lagrangian constrained by $p$-wave $\pi\pi$ measurements and the pion electromagnetic form factor\,\cite{Rapp:1999us}. In the other case, the in-medium $\rho$-meson propagator is calculated based on the hadronic many-body theory, leading to a broadened $\rho$ spectral function mainly due to the interactions of the $\rho$ with (anti)baryons. Predictions from the microscopic dynamic transport model PHSD\,\cite{Linnyk:2011vx} (see subsection \ref{models}) are also shown. In the hadronic phase, the dielectron production is calculated using an in-medium modified $\rho$ spectral function which changes dynamically during the propagation through the medium and evolves towards the $\rho$ spectral functions in vacuum. On the one hand, the vacuum $\rho$ plus QGP scenario in Rapp's implementation does not describe the data well. On the other hand, the calculations from Rapp and PHSD including a broadening of the $\rho$ spectral function show a reasonable agreement with the measured excess yield. The same models can reproduce the low-mass dilepton enhancement measured at SPS energy\,\cite{ CERES:2006wcq, NA60:2006ymb}. The hadronic medium at top RHIC energy is thus similar to that created at SPS energy, even if the lifetime of the fireball and the total thermal radiation yield, in particular from the partonic phase, are larger.\\

In the right panel of \textbf{Figure \ref{secondfigure}} the centrality dependence of the excess yield in the mass range $0.3 < \mee < 0.76$\,\GeVmass ~is shown\,\cite{STAR:2015tnn}. The integrated excess yield scaled by the mean number of participating nucleons ($N_{\rm part}$) in Au--Au collisions is plotted as a function of the centrality ($N_{\rm part}$), together with the dielectron yields in the $\rho$ (A: $0.3-0.76$\,\GeVmass), $\omega$ (B: $0.76-0.80$\,\GeVmass) and $\phi$ (C:$0.98-1.05$\,\GeVmass) mass regions. For the two last sets (B and C), the yields were also scaled by $N_{\rm part}$ but without subtracting the cocktail before. Whereas the $\omega$-like and $\phi$-like dielectron yields exhibit an $N_{\rm part}$ scaling, the excess yield in region A increases faster than $N_{\rm part}$. The result of a power-law fit ($\propto N_{\rm part}^{\rm a}$) to the excess points is shown by the dashed curve with $a = 0.44\pm 0.10$ (stat. + uncorrelated sys.). Such centrality dependence is expected from theoretical calculations for thermal radiation including in-medium modifications of the $\rho$ spectral function, although the transport model (PHSD\,\cite{Linnyk:2011vx}) seems to predict a weaker centrality dependence than the fireball model of Rapp\,\cite{Rapp:2013nxa}. The slope from Rapp's calculations is in fair agreement with that derived from the data, even if it slightly overestimates the centrality dependence of the excess. On the other hand, the reported centrality dependence of the PHSD model exhibits some tension with the data.

\subsection{Beam energy scan at RHIC: 7.7-62.4\,GeV}

Starting in 2010, the STAR experiment launched a Beam Energy Scan (BES) program for a detailed investigation of the QCD phase diagram at different temperature and $\mu_{\rm B}$. With the data collected during the first phase of this program (BES-I), dielectron production was studied in Au--Au collisions at $\snn = $19.6, 27, 39 and 62.4\,GeV\,\cite{STAR:2015zal, STAR:2023wta}. At the lowest collision energies, only a limited amount of data could be recorded due to the detector and accelerator capabilities. Further dielectron measurements at $\snn = $7.7, 9.2, 11.5, 14.6, 19.6\,GeV\,\cite{Han:2024nzr,sQM2024starprel,HP2024starprel} and 27, 54.4\,GeV\,\cite{STAR:2024bpc} in Au--Au collisions, as well as in Ru--Ru and Zr--Zr collisions at $\snn =$\,0.2,TeV\,\cite{HP2024starprelbis}, were possible with the BES-II, after an increase of the luminosity delivered by the accelerator and STAR detector upgrades. From the measured yields, \pt\,spectra and particle ratios of identified hadrons ($\pi^{\pm}$, K$^{\pm}$, p and $\rm \bar{p}$), the temperatures at chemical ($T_{\rm ch}$) and kinematic ($T_{\rm kin}$) freeze-outs were shown to be approximately constant in the \snn\,\,range 19.6-200\,GeV for minimum bias Au--Au collisions, whereas they decrease significantly for smaller \snn\,\cite{STAR:2017sal}. Similarly, the total baryon density (at chemical freeze-out) increases substantially with decreasing \snn\,\,below 19.6\,GeV, while it has no significant \snn-dependence above. Note that $T_{\rm ch}$ was found to be consistent with the phase transition temperature $T_{\rm c}$ from lQCD for $\snn \geq 11.5$\,GeV\,\cite{Braun-Munzinger:2003htr}. Dileptons have the potential to shed light on the temperature of the medium at earlier stages.\\

\begin{figure}[h]
  \begin{minipage}{.5\textwidth}
  \centering
  
  \includegraphics[width=0.68\linewidth,angle=-90]{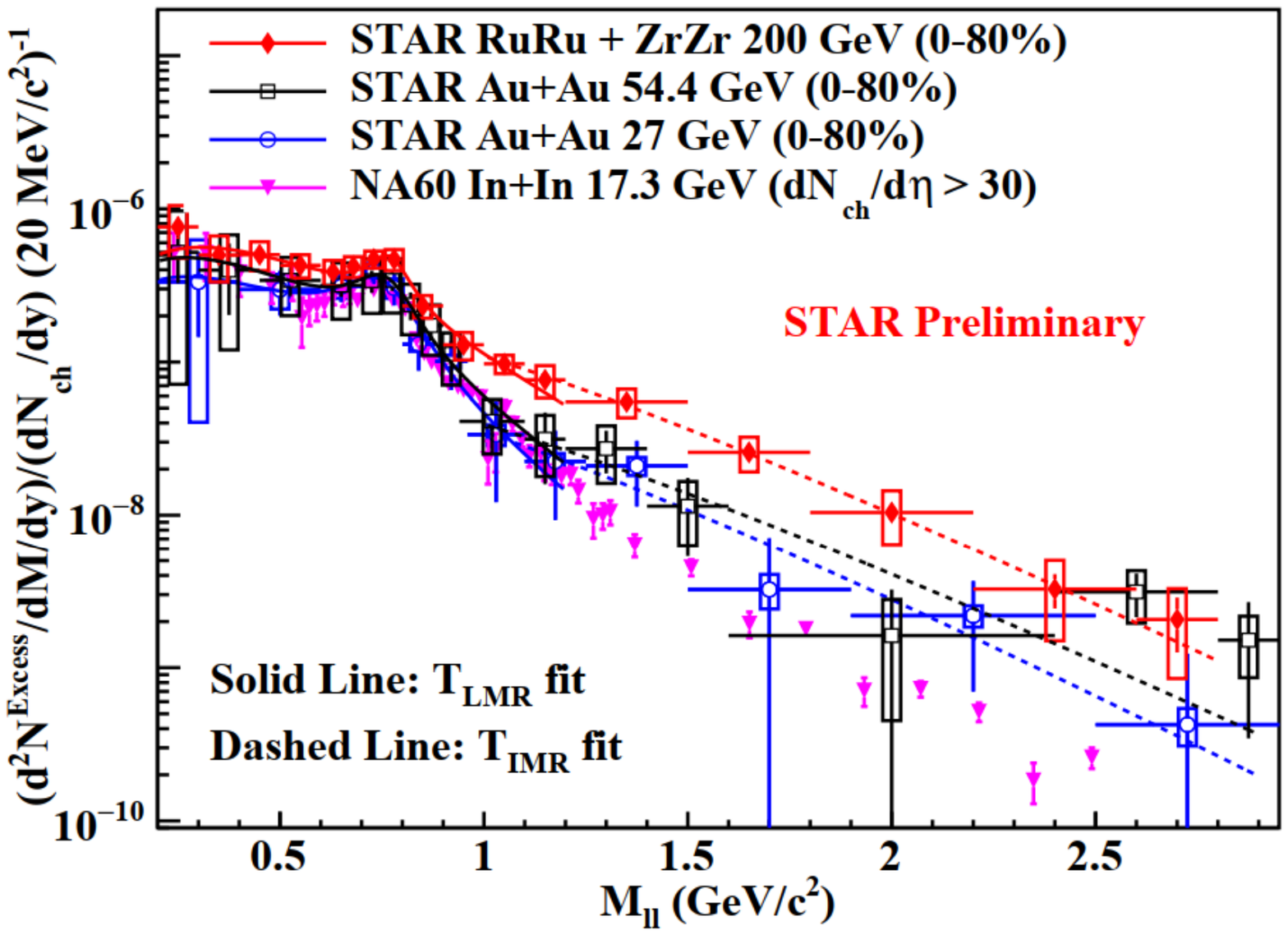}
\end{minipage}%
\begin{minipage}{.5\textwidth}
  \centering
  \includegraphics[width=0.74\linewidth,angle=-90]{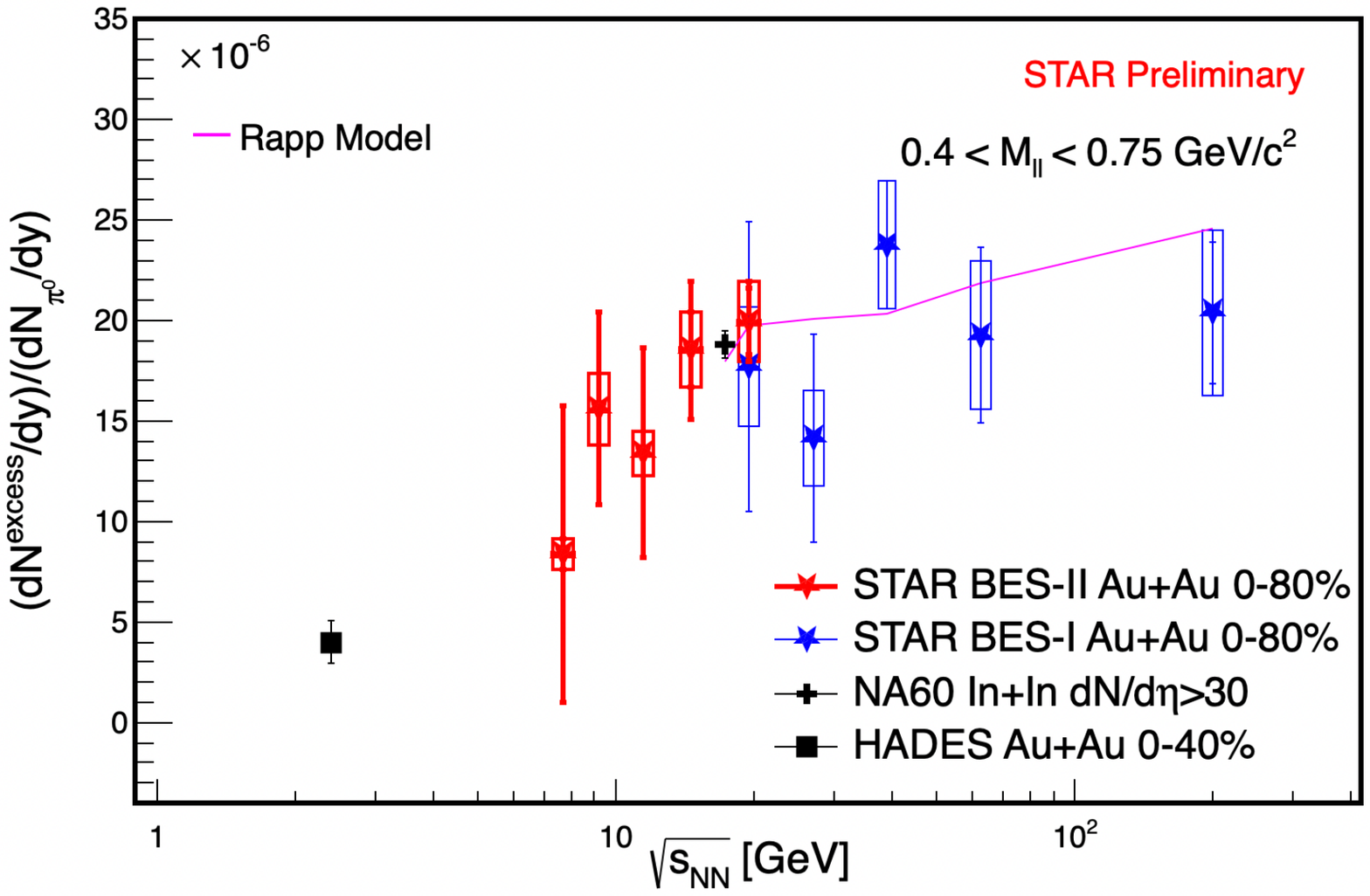}
\end{minipage}
\caption{Left panel: Acceptance-corrected dilepton excess mass spectra normalised by the charged-particle multiplicity in different colliding systems and at different $\snn$ from STAR\,\cite{STAR:2024bpc, HP2024starprelbis} and NA60\,\cite{NA60:2008ctj}, fitted to extract temperature parameters. Vertical bars and boxes around data points represent the statistical and systematic uncertainties, respectively. Panel from Reference\,\cite{HP2024starprelbis}. Right panel: Collision-energy dependence of the integrated dilepton excess yields in the mass range $0.4 < \mll < 0.75$\,\GeVmass\,\,measured by STAR\,\cite{STAR:2015zal, STAR:2023wta, Han:2024nzr, sQM2024starprel, HP2024starprel, STAR:2024bpc}, NA60\,\cite{NA60:2008ctj} and HADES\,\cite{HADES:2019auv}, normalised by the neutral-pion multiplicity and compared to calculations\,\cite{Rapp:2013nxa}. Panel from Reference\,\cite{HP2024starprel}.\label{thirdfigure}}

\end{figure}

The STAR Collaboration reported significant excess yields of dielectrons in the low-mass region ($0.4 < \mee < 0.75$\,\GeVmass) compared to the expected contributions from hadronic decays in mimimum bias Au--Au collisions for all \snn\,\, under consideration\,\cite{STAR:2015zal, STAR:2023wta, Han:2024nzr, sQM2024starprel,HP2024starprel, STAR:2024bpc}. At higher mass, where correlated open heavy-flavour hadron decays ($\snn \geq 27$\,\GeV) or the Drell-Yan process ($\snn < 19.6$\,GeV) are the main source of $\rm e^{+}e^{-}$ pairs, the measured dielectron spectra are slightly above the cocktails only for $\snn \geq 19.6$\,GeV. After subtraction of the hadronic cocktails excluding the $\rho$ contribution, the excess spectra were corrected for detector acceptance estimated with Monte Carlo simulations taking as inputs some assumed virtual photon spectra. Examples of excess yields are shown in the left panel of \textbf{Figure \ref{thirdfigure}}. The spectra are normalised to the measured charged-particle multiplicity at midrapidity (${\rm d}N_{\rm ch}/{\rm d}y$) to cancel the volume effect and compare the measurements among different colliding species and beam energies.


At low mass ($\mll < 0.8$\,\GeVmass), the normalised excess distributions are similar within uncertainties for $\snn \geq $17.3\,GeV, although results in minimum bias Ru--Ru $+$ Zr--Zr collisions tend to be higher\,\cite{HP2024starprelbis}. The normalized excess yields integrated in the mass range $0.4 < \mll < 0.75$\,\GeVmass ~are depicted as a function of \snn\,\,in the right panel of \textbf{Figure \ref{thirdfigure}}. In this mass range, the excess yield in minimum bias Au--Au collisions for $20 < \snn < 200$\,GeV is about a factor 5 larger than the expected dielectron production from $\rho$ decays at chemical freeze-out\,\cite{STAR:2024bpc}. Calculations for thermal radiation from the hadronic and partonic phases with in-medium modified $\rho$ spectral function reproduce fairly well the data\,\cite{Rapp:2013nxa}. The same predictions describe the excess mass spectra up to a mass of about 1.6\,GeV/$c^{\rm 2}$, although the data suffer from relatively large statistical uncertainties above 1.2\,GeV/$c^{\rm 2}$\,\cite{STAR:2024bpc}. While for $\snn > 20$\,GeV the normalised low-mass excess yields do not exhibit any clear $\snn$-dependence, at lower $\snn$ a hint for a decrease may be observed. However, the results suffer from large statistical uncertainties. Fixed target experiments (CBM at FAIR\,\cite{CBM:2016kpk} or NA60+ at the CERN SPS\,\cite{NA60:2022sze}) are designed to resolve the gap between $\snn = 17.3$\,GeV (NA60\,\cite{NA60:2008ctj}) and $\snn = 2.42$\,GeV (HADES\,\cite{HADES:2019auv}) with high accuracy. It is worth to notice that the excess yield at low mass ($0.3 < \mll < 0.7$\,\GeVmass) was proposed as a chronometer of the total lifetime of the fireball\,\cite{Rapp:2014hha}. However, the mentioned mass window was explicitly excluding the $\rho$ pole mass and extended towards lower mass to include a significant contribution from the QGP.

At higher mass ($\mll > 1$\,\GeVmass), the excess of $\rm l^{+}l^{-}$ pairs is expected to originate predominantly from the partonic phase. Predictions for QGP thermal radiation with production rates computed up to NLO at finite $\mu_{\rm B}$ and integrated over space-time using realistic hydrodynamics describe fairly well the excess spectra measured in minimum bias Au--Au collisions at $\snn \geq 19.6, 27, 39, 62.4, 200$\,GeV\,\cite{Churchill:2023zkk}. Effective temperatures $T_{\rm eff}$ can be extracted from the inverse slope of the predicted QGP thermal dilepton invariant mass spectra ($\propto \mll^{3/2}*{\rm e^{-\mll/T_{\rm eff}}}$). The calculations show that there is a linear relationship between $T_{\rm eff}$ and the average temperature of the fluid used at the beginning of the QGP fireball expansion in the simulated collisions. In this sense, effective temperatures derived from measured dilepton spectra have the potential to provide information on the early temperature of the fireball. Such measurements were already suggested in Ref.\,\cite{Rapp:2014hha}. In the left panel of \textbf{Figure \ref{thirdfigure}} the measured excess spectra were fitted accordingly in the $1 < \mll < 2.9$\,\GeVmass\,\,range (dashed lines) to determine $T^{\rm data}_{\rm eff} = T_{\rm IMR}$. The obtained values of $T_{\rm IMR}$ are high, well above $T_{\rm c}$, higher than those in the calculations\,\cite{Churchill:2023zkk,STAR:2024bpc}. Contributions from the pre-equilibrium phase, for which predictions are not yet fully available at RHIC energies, may play a role in the understanding of $T_{\rm IMR}$. Moreover, the unknown modification of the correlated heavy-flavour background in heavy-ion collisions and the lack of precise Drell-Yan calculations at low mass limit the estimation of the excess yield and may not be fully reflected in the quoted systematic uncertainties. The STAR collaboration extended the temperature extraction method towards lower masses ($T_{\rm LMR}$ for $0.2 < \mll < 1.2$\,\GeVmass), using a fit function that combines the in-medium resonance structure and the continuum thermal distribution. Although the validity of such an approach has to be demonstrated, the extracted temperatures $T_{\rm LMR}$ are similar for minimum bias Au--Au collisions at $\snn =$27 and 54.4\,GeV, as well as in In--In collisions at $\snn =$17.3\,GeV, and consistent with $T_{\rm c}$. This suggests that the low-mass thermal dileptons are predominantly emitted over an extended period of time at high density around a fixed temperature.



\section{Results from the LHC} 

Pb--Pb collisions at the LHC provide the hottest and longest-lived QGP at $\mu_{\rm B} \simeq 0$, close to the conditions in the early universe and where lattice QCD calculations can be performed. Among the different LHC experiments, ALICE is so far the only one that offers the possibility to track and identify leptons at low momenta in an environment of high charged-particle multiplicity. Due to the large magnetic fields used in their detector set-up (3.8 and 2\,T, respectively) and the amount of absorber material, the acceptances of the CMS and ATLAS experiments for leptons is restricted to relatively large \pt, i.e. $\pt \geq 2.5$\,GeV or higher depending on the lepton species ($\rm e^{\pm}$ or $\mu^{\pm}$) and experiment. In addition, the LHCb Collaboration has published results on $\rm \mu^{+}\mu^{-}$ pairs produced at forward rapidity $2 \leq \eta \leq 4.5$ in pp collisions at $\snn = 13.6$\,TeV for $\pt \geq 0.5$\,GeV and $p \geq 10$\,GeV\,\cite{LHCb:2017trq}. Upgrades of the LHCb detector are planned for the LHC Run 5 (starting in 2036) to allow such measurements also in colliding systems with high charged-particle multiplicity\,\cite{LHCb:2018roe}. Low-mass lepton measurements in central Pb--Pb collisions are, however, not yet possible with the current LHCb detector setup. Thus, the first results on low-mass dielectron production in Pb--Pb collisions at the LHC were provided by the ALICE Collaboration with the data collected during the LHC Run 1 at \twosevensixnn\,\cite{ALICE:2018ael} in 2011 and the LHC Run 2 at \fivenn\,\cite{ALICE:2023jef} in 2018. The ALICE electron detection system at midrapidity consists, among other subdetectors, of a large TPC, a TOF detector, and an Inner Tracking System (ITS) with full azimuthal coverage installed in a central barrel at midrapidity ($\eta < 0.9$). The high uniformity of the azimuthal acceptance of the ALICE setup allows a very precise estimation of the combinatorial and correlated backgrounds using the same-event like-sign method. On the other hand, due to the larger charged-particle multiplicity, the signal-to-background ratio is up to an order of magnitude smaller than at RHIC for similar fiducial cuts ($\pt > 0.2$\,GeV and $\eta < 0.8$). 
The combinatorial background scales approximately with $({\rm d}N_{\rm ch}/{\rm d}y)^{\rm 2}$, while the dielectron yield itself is roughly proportional to  ${\rm d}N_{\rm ch}/{\rm d}y$.\\

\begin{figure}[h]
\begin{minipage}{.5\textwidth}
  \centering
  \includegraphics[width=1.\linewidth]{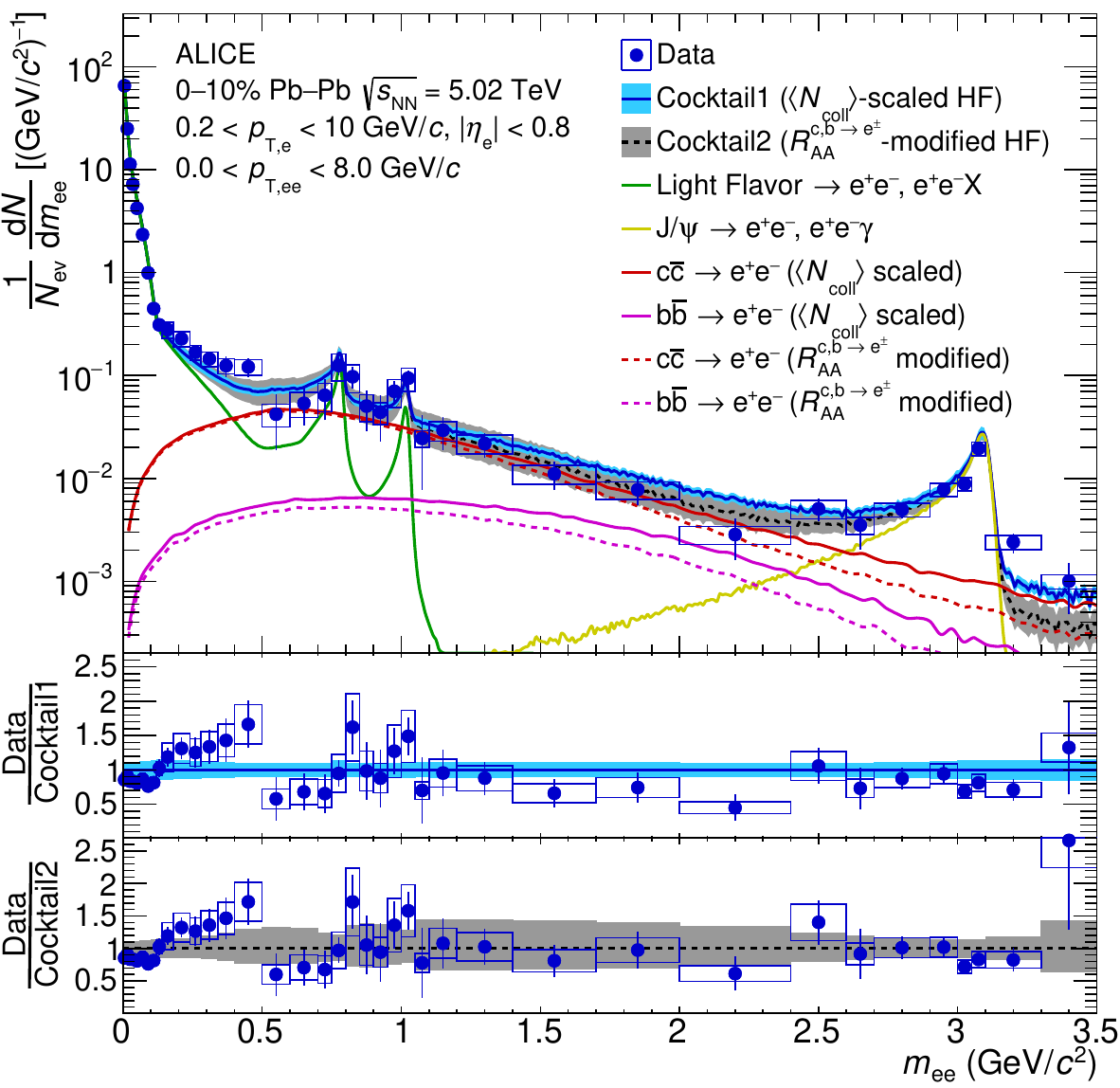}
\end{minipage}%
\begin{minipage}{.5\textwidth}
  \centering
  \includegraphics[width=1.\linewidth]{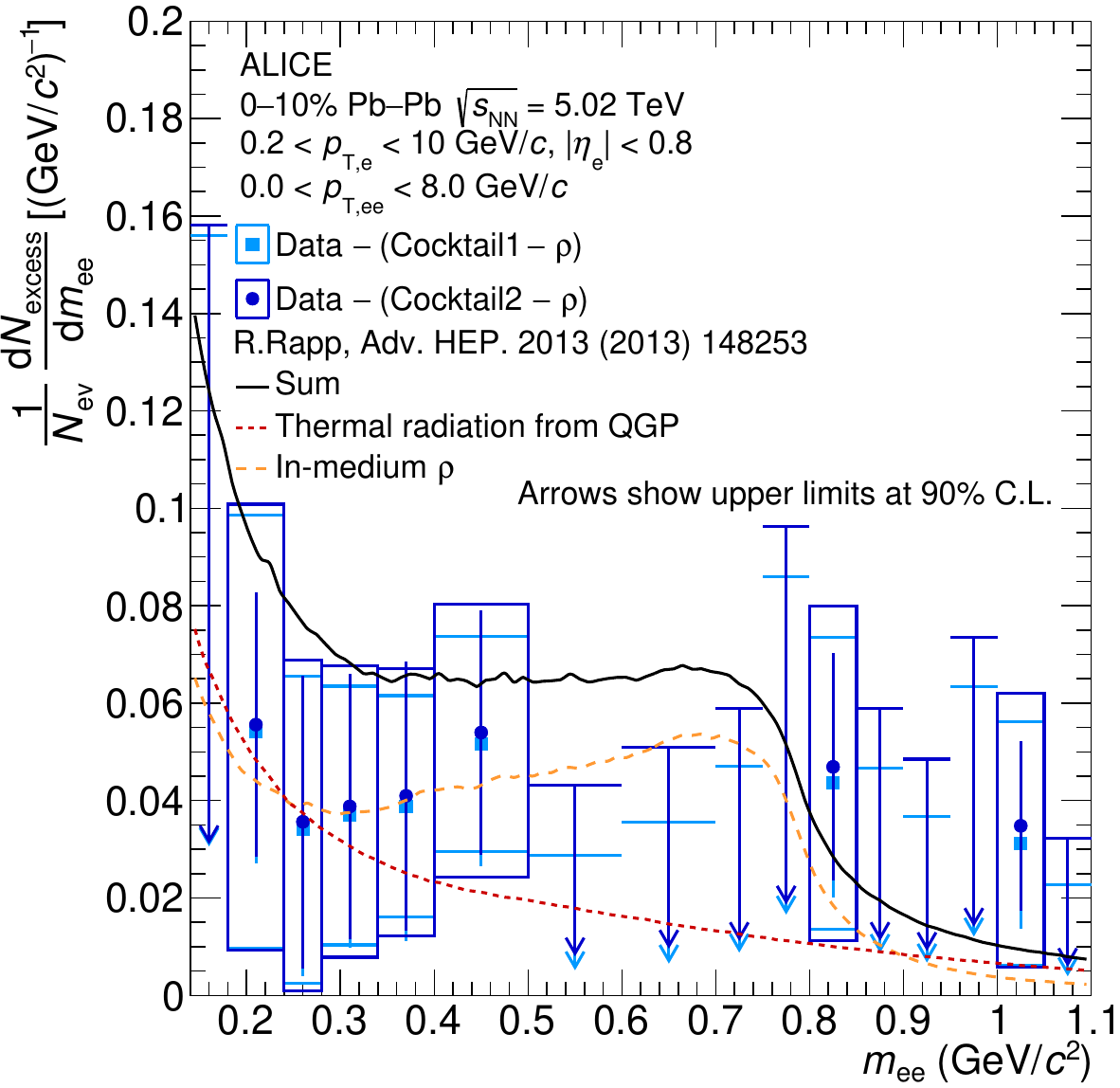}
\end{minipage}
\caption{Left panel: Dielectron \mee-differential yield in the 10\% most central Pb--Pb collisions at \fivenn, compared with the expected $\rm e^{+}e^{-}$ contributions from known hadronic decays (cocktails), including two different estimations for dielectrons from correlated heavy-flavour hadron decays. The error bars and boxes represent the statistical and systematic uncertainties of the data, whereas the bands show the uncertainties of the hadronic cocktails. Panel from Reference\,\cite{ALICE:2023jef}. Right panel: Corresponding excess yield of dielectrons with respect to the cocktails, compared with predictions from the model of Rapp\,\cite{Rapp:2013nxa}. The error bars and boxes represent the total statistical and systematic uncertainties including those from the cocktails\,\cite{ALICE:2023jef}. Panel from Reference\,\cite{ALICE:2023jef}. \label{fig:4}}
\end{figure}

In the left panel of \textbf{Figure \ref{fig:4}} the dielectron yield measured in the ALICE acceptance ($0.2 < \pt < 10$\,\GeVc\,\, and $|\eta| < 0.8$) in the 0--10\% most central Pb--Pb collisions at \fivenn\,\, is shown as a function of the invariant mass. The data are compared to the expected contributions from known hadronic sources neglecting any medium effect for the $\rho$ mesons. The large background from correlated heavy-flavour hadron decays is estimated from dielectron measurements in pp collisions\,\cite{ALICE:2020mfy} leaving out any cold or hot medium effects (Cocktail 1) or calculated in a somewhat model-dependent way using information from single heavy-flavour decay electron measurements\,\cite{ALICE:2019nuy} and computations performed with the EPS09 nuclear parton distribution functions\,\cite{Eskola:2009uj} (Cocktail 2). Incorporating heavy-flavour modifications improves the description of the data by the sum of the background contributions in the mass range where open-charm and beauty hadron decays are predicted to dominate the dielectron yield ($1.2 < \mee < 2.6$\,\GeVmass). However, it increases substantially the systematic uncertainties. No clear excess over the cocktails is observed, although a hint for an enhanced dielectron production on the 1$\sigma$ level is indicated in $0.18 < \mee < 0.5$\,\GeVmass.

\begin{figure}[h]
\begin{minipage}{0.5\textwidth}
  \centering
  \includegraphics[width=1.\linewidth]{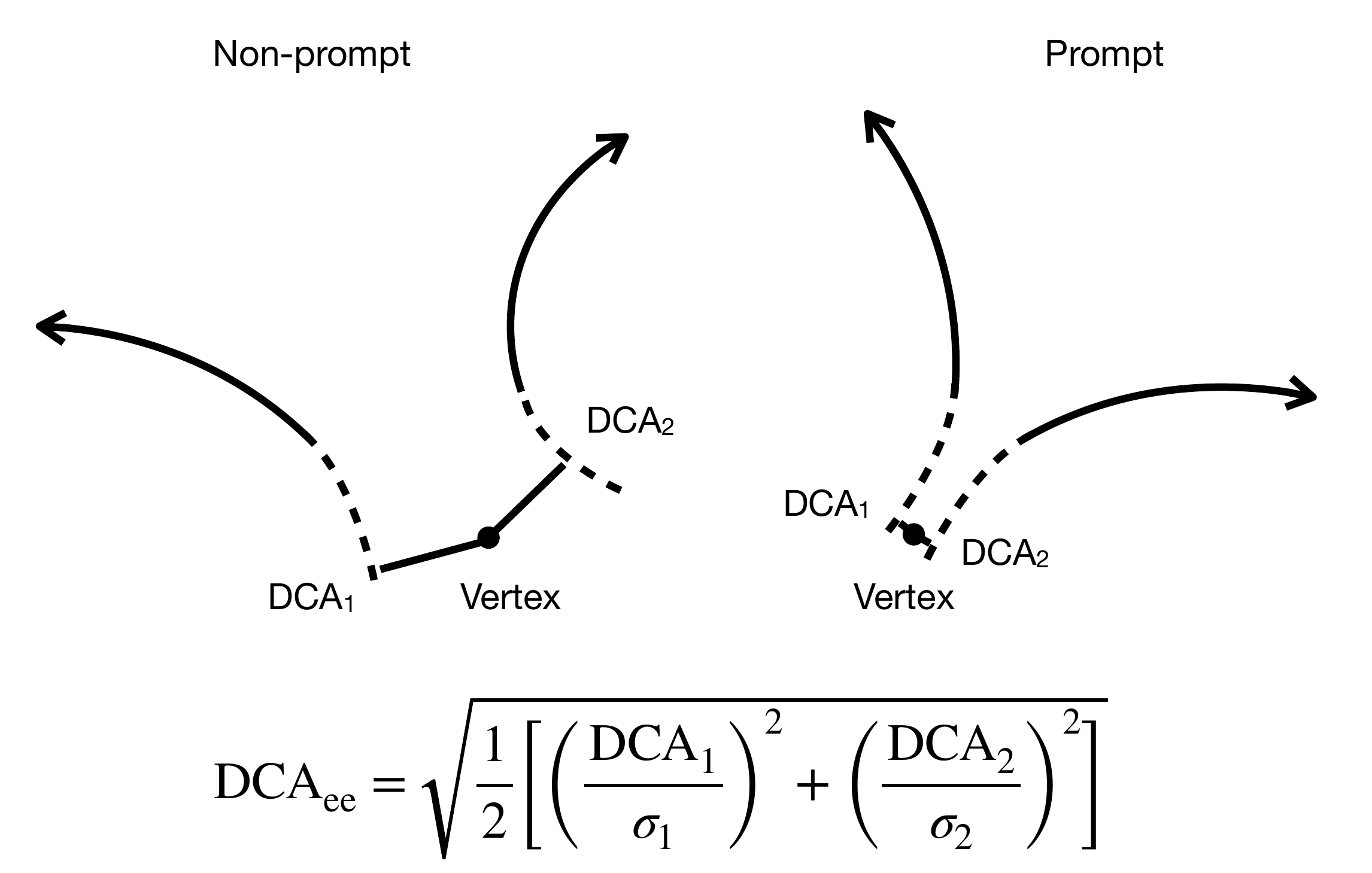}
\end{minipage}%
\begin{minipage}{0.5\textwidth}
  \centering
  \includegraphics[width=1.\linewidth]{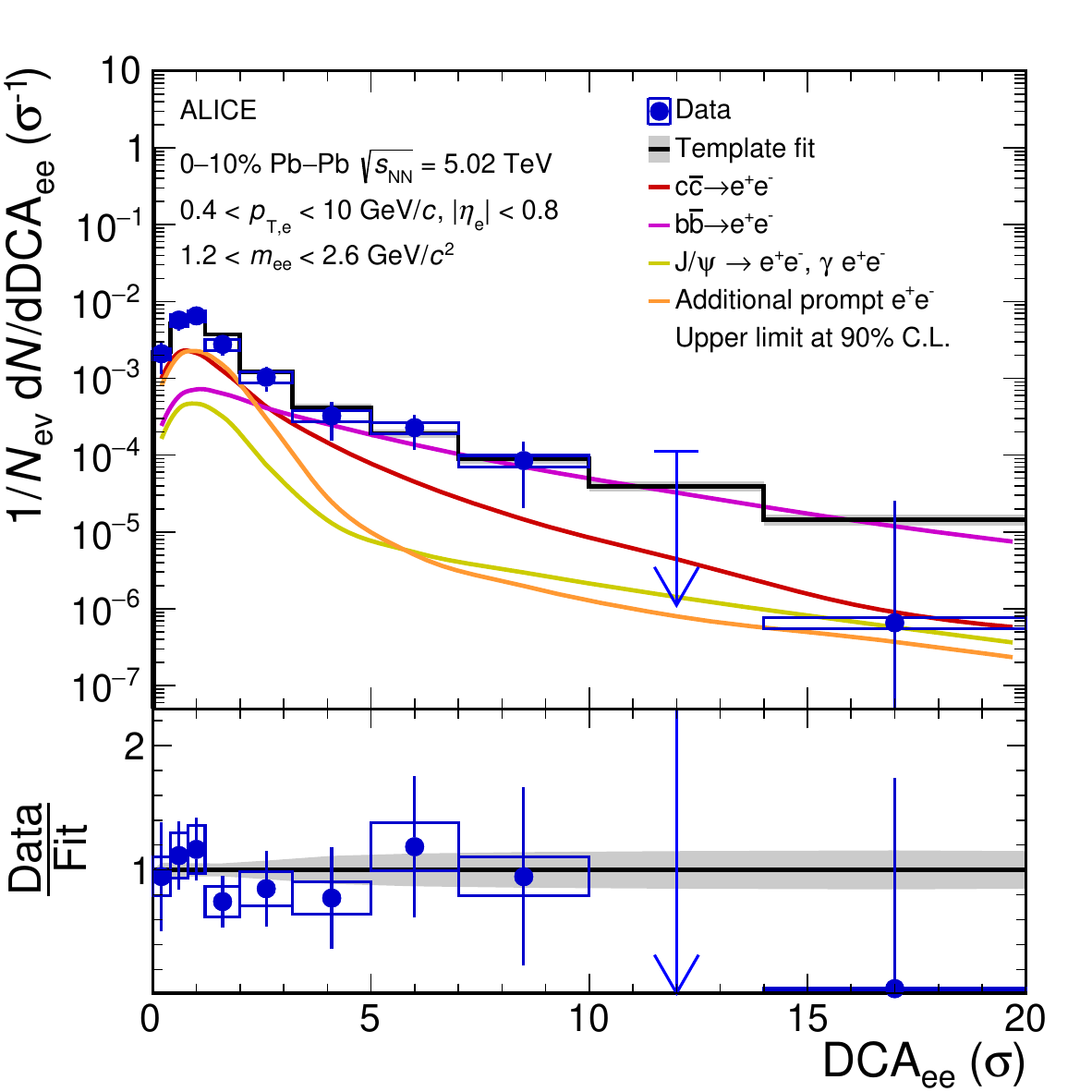}
\end{minipage}
\caption{Left panel: Sketch of the dielectron spatial topology with the definition of the distance-of-closest approach (DCA) to the reconstructed collision vertex of the single $\rm e^{\pm}$ and $\rm e^{+}e^{-}$ pairs. Panel provided by Sebastian Scheid. Right panel: Fit of the inclusive $\rm e^{+}e^{-}$ yield in the 10\% most central Pb--Pb collisions at \fivenn\,\,as a function of DCA$_{\rm ee}$ in the mass range $1.2 < \mee < 2.6$\,\GeVmass. The error bars and boxes represent the statistical and systematic uncertainties of the data. Panel from Reference\,\cite{ALICE:2023jef}.\label{fig:5}}
\end{figure}

In the right panel of \textbf{Figure \ref{fig:4}} the cocktails, excluding the contribution from $\rho$ mesons, are subtracted from the data to obtain the excess spectrum in the mass range where an excess is observed at RHIC energies. The excess yield is consistent with calculations for  thermal radiation from the hadronic and partonic phase (sum of orange and red lines), although a tension of the order of 4$\sigma$ is observed for $0.5 < \mee < 0.8$\,\GeVmass. Both, the uncertainties of the data and of the cocktails need to be reduced. For the latter, an elegant but challenging way to determine and suppress experimentally the huge physical background from correlated heavy-flavour hadron decays is to make use of the finite decay length of open-beauty ($c\tau \simeq 450$\,$\mu$m) and open-charm ($c\tau \simeq 150$\,$\mu$m) hadrons. This requires an inner tracking system with high-precision vertexing capability to measure the displaced vertices of heavy-flavour hadron decays. As shown in the left panel of \textbf{Figure \ref{fig:5}}, leptons from non-prompt dilepton sources, like heavy-flavour hadron decays, have on average a larger distance-to-closest approach to the reconstructed collision vertex (DCA$_{\rm 1}$ or DCA$_{\rm 2}$) than those originating from strong and electromagnetic decays or processes happening at the collision point. By defining a pair DCA variable (DCA$_{\rm ee}$) based on the DCA of the single leptons, prompt $\rm l^{+}l^{-}$ pairs from light-meson decays, the Drell-Yan process or thermal radiation in the medium can be disentangled from non-prompt $\rm l^{+}l^{-}$ pairs from heavy-flavour hadron decays. To account for the \pt ~dependence of the DCA detector resolution for prompt dileptons, it is advantageous to normalise the single DCAs by their resolution ($\sigma_{\rm 1}$ or $\sigma_{\rm 2}$), which is estimated from the covariance matrix of the track reconstruction parameters, as demonstrated by the NA60 collaboration\,\cite{NA60:2008dcb}.

In the right panel of \textbf{Figure \ref{fig:5}} the measured DCA$_{\rm ee}$ distribution of $\rm e^{+}e^{-}$ pairs in central Pb--Pb collisions at \five ~is shown in $1.2 < \mee < 2.6$\,\GeVmass for $0.4 < \pt < 10$\,\GeVc\,\, and $|\eta| < 0.8$. In this case, the DCA was calculated in the plane transverse to the beam axis. The data are fitted with template distributions from Monte Carlo simulations. The fit includes templates for open-charm (red line) and open-beauty (magenta line) $\rm e^{+}e^{-}$ pairs and a prompt dielectron contribution (orange line). Even if the interpretation of the results is limited by large statistical uncertainties, the data are found to be consistent with a large suppression of dielectrons from $\rm c\bar{c}$ compared to the extracted corresponding yield in pp collisions scaled with the number of binary nucleon–nucleon collisions. A similar result is obtained for $\rm e^{+}e^{-}$ pairs from $\rm b\bar{b}$, although less pronounced. Finally, the data appear to be compatible with an additional prompt component, presumably from thermal radiation. This is the first experimental sign of thermal radiation from the QGP in Pb--Pb collisions at the LHC, albeit with a significance of only 1$\sigma$.


\section{Perspectives with recent upgrades} 

During the LHC Long Shutdown 2 (LS2) from 2019 to 2021, upgrades of the LHC were performed to increase the Pb--Pb luminosity, reaching an interaction rate of about 50\,kHz\,\cite{Arduini:2024kdp}. In parallel, a major upgrade of the ALICE detector was completed, with a focus on improving its event rate capabilities and significantly enhancing the tracking and vertexing capabilities at low momentum\,\cite{ALICE:2012dtf,ALICE:2023udb}. In particular, the new readout system of the TPC based on GEMs (Gas Electron Multipliers) allows continuous readout of the data, increasing the rate by about a factor 1000 in pp collisions and up to 100 in Pb--Pb collisions as compared to Run 2\,\cite{Lippmann:2014lay,TheALICECollaboration:2015xke,ALICETPC:2020ann}. In total, ALICE aims for recording an integrated luminosity of 13\,nb$^{-1}$ in Pb--Pb collisions at \fivennnew\,\,during the LHC Runs 3 (2022-2026) and 4 (2030-2033), which is about two orders of magnitude above what was published until now for dielectrons. At the same time, the vertex pointing resolution of the detector is at least three times better than during the LHC Runs 1 and 2, owing to the new inner tracking system ITS2 based solely on monolithic active pixel sensors with smaller pixels and a first layer closer to the interaction point\,\cite{ALICE:2013nwm}. Further improvements of the ALICE vertexing capabilities are planned starting from 2030 by replacing the three innermost ITS layers during the LS3\,\cite{ITS3,ITS3bis}. The usage of the latest CMOS MAPS, i.e.\,curved wafer-scale ultra-thin ($< 50$\,$\mu$m) silicon sensors without the need of massive support structures should result in a DCA resolution that is two times better than with the present ITS2. This will allow a still better separation of the prompt radiation emitted by the hot medium in heavy-ion collisions and the correlated heavy-flavour hadron decay background. A reduction of the combinatorial background is a priori also expected due to the progressive decrease of detector material close to the interaction point. Moreover, according to simulations\,\cite{ALICE:2012dtf,Lippmann:2014lay,ALICE:2013nwm,ITS3,Citron:2018lsq}, further improvement of the signal-to-background ratio will be possible thanks to the improved efficiency of pre-filtering the reconstructed tracks originating from $\pi^{0}$ decays and $\gamma$ conversions in the detector material, as already tested with the ALICE Run 1 pp data\,\cite{ALICE:2018fvj}. Therefore, first significant measurements of thermal radiation from the QGP and the $\rho$ spectral function in Pb--Pb collisions at \fivennnew\,\,are in reach by the end of the LHC Run 4. Note that the installation of a muon forward tracker\,\cite{mft} in front of the absorber of the ALICE Muon Spectrometer during the LS2 allows an extension of the prompt dilepton analyses to forward rapidities ($-3.6 < \eta < -2.45$) with dimuons.

\begin{figure}[h]
\begin{minipage}{0.5\textwidth}
  \centering
  \includegraphics[width=1.\linewidth]{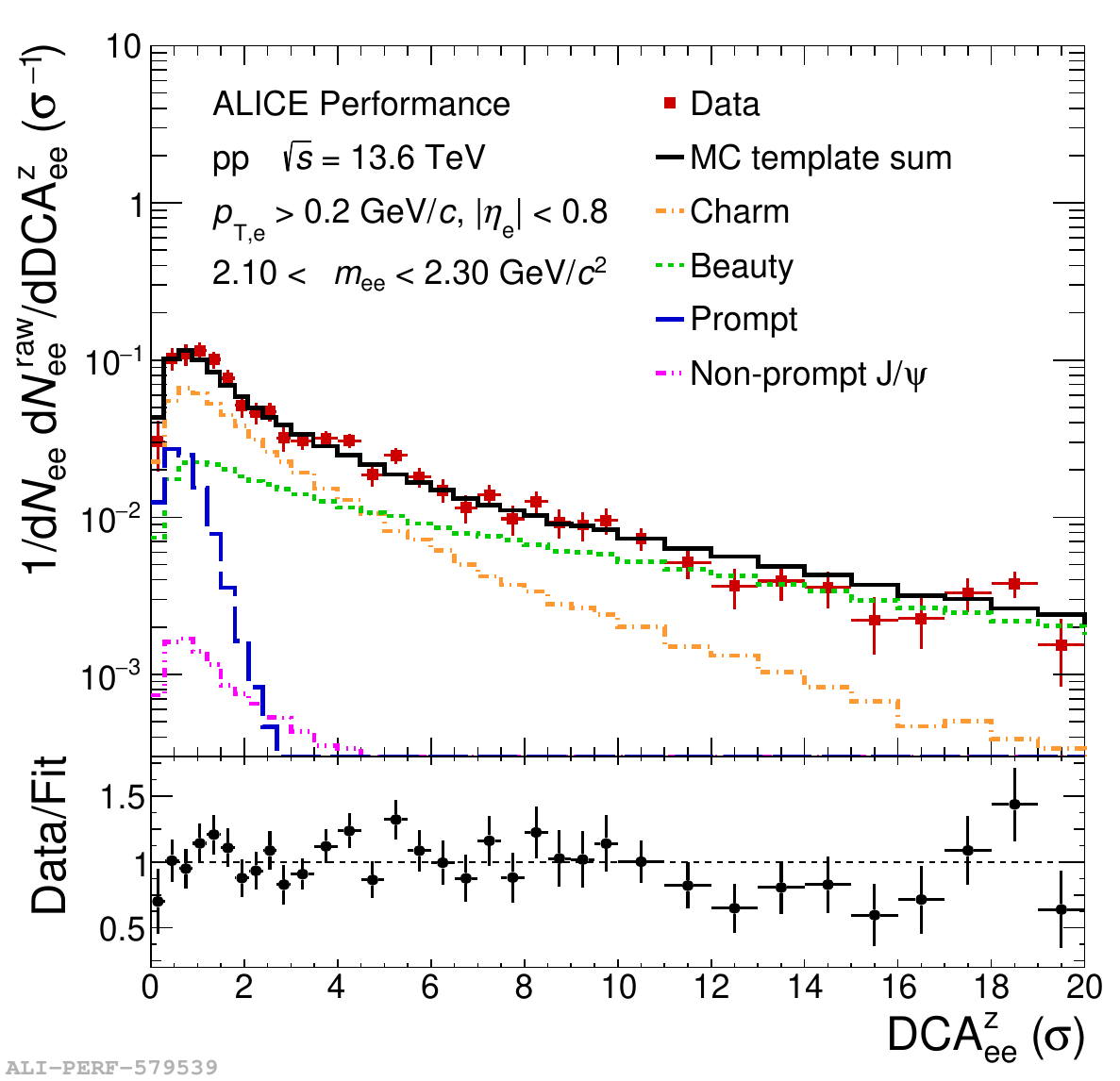}
\end{minipage}%
\begin{minipage}{0.5\textwidth}
  \centering
  \includegraphics[width=1.\linewidth]{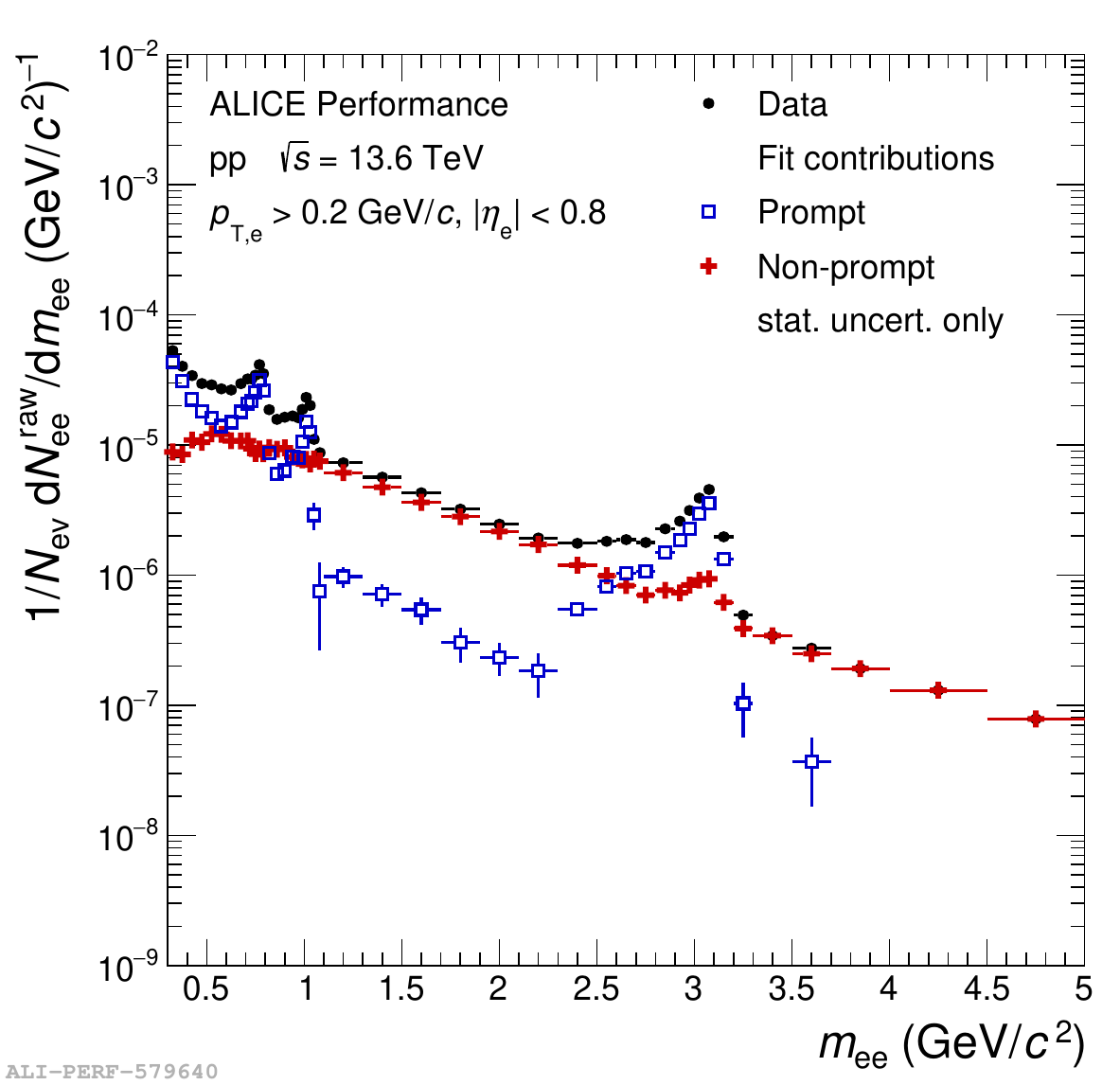} 
\end{minipage}
\caption{Left panel: Fit of the inclusive raw $\rm e^{+}e^{-}$ yield in pp collisions at \thirdteennew\,\,as a function of DCA$^{\rm z}_{\rm ee}$ in the mass range $2.1 < \mee < 2.3$\,\GeVmass. Panel from Reference\,\cite{JeromeJungHardProbes2024}. Right panel: Raw mass spectra of inclusive, prompt, and non-prompt $\rm e^{+}e^{-}$ pairs in pp collisions at \thirdteennew. Only statistical uncertainties are shown. Panel from Reference\,\cite{JeromeJungHardProbes2024}\label{fig:6}}
\end{figure}

A data sample of about 3\,nb$^{-1}$ Pb--Pb collisions was already accumulated in the course of the LHC 2023 and 2024 heavy-ion campaigns with a data acquisition rate reaching 50\,kHz. While the calibration of these data is still ongoing, the improved detector performance is already evident in the latest status report on the analysis of the minimum bias pp collisions at \thirdteennew\,\, recorded in 2022, corresponding to an integrated luminosity of 0.97\,pb$^{-1}$. For comparison, ALICE reported an integrated luminosity of 0.03\,pb$^{-1}$ for the full LHC Run 2 period. In the left panel of \textbf{Figure \ref{fig:6}} the measured DCA$^{\rm z}_{\rm ee}$ distribution of $\rm e^{+}e^{-}$ pairs with $2.1 < \mee < 2.3$\,\GeVmass\,\,in pp collisions is fitted with the sum of templates obtained from MC simulations for each dielectron source.\begin{marginnote}[]
\entry{DCA$^{\rm z}_{\rm ee}$}{Pair DCA based on single electron DCA measured in the beam axis direction.}
\end{marginnote} The data are not corrected for efficiency but no significant changes to the templates are expected after correction. Owing to the normalization of DCA$^{\rm z}_{\rm ee}$ to the detector resolution, the prompt template does not depend on the detector layout. Comparing the shapes of the DCA$_{\rm ee}$ templates for dielectrons from heavy-flavour hadron decays with the ones of \textbf{Figure \ref{fig:5}} one can see that their average DCA$_{\rm ee}$ is shifted towards larger values for the Run 3 data compared to the Run 2 results. This implies a better separation of prompt and non-prompt dielectron sources. In the right panel of \textbf{Figure \ref{fig:6}} the prompt and non-prompt contributions to the raw dielectron mass spectrum are unfolded based on such DCA$_{\rm ee}$ fits performed in \mee\,\,slices. Whereas the continuum from correlated heavy-flavour hadron decays and the contribution from non-prompt $J/\psi$ decays are present in the non-prompt distribution, peaks from light vector meson decays ($\omega$, $\rho$, $\phi$) and prompt $J/\psi$ are well visible in the prompt spectrum. This represents an important milestone on the way to measuring prompt radiation in heavy-ion collisions.


\section{Next-generation dilepton spectrometers - ALICE 3 and LHCb-UII}

The ALICE and LHCb Collaborations are both planning further major upgrades to their detectors during the LS4 (2034-2035). \\

The LHCb Phase II upgrade program (LHCb-UII) \,\cite{LHCb:2018roe} aims at exploiting the full pp luminosity of the High-Luminosity LHC in Run 5 while maintaining the detector performance as in Runs 3 and 4. LHCb-UII will also allow measurements in central heavy-ion collisions while in Run 4 the most central Pb--Pb collisions will not be accessible with the existing LHCb tracking detectors. A new VErtex LOcator (VELO) detector including time information is foreseen together with new tracking systems upstream (Upstream Tracker) and downstream (Mighty Tracker) of the LHCb bending magnet as well as further upgrades of the detectors for particle identification (RICH, TOF, ECal and muon stations). Altogether, the LHCb-UII upgrade is expected to enable the study of low-mass dilepton production in the rapidity range $2 < y < 5$ with dimuons and dielectrons in Run 5, where for electrons the tracking efficiency at low \pt\,is presumably reduced due to radiative energy loss before the LHCb dipole magnet deflecting them outside of the downstream tracker acceptance\,\cite{LHCb:2019gvd}.\\

In 2022 the ALICE collaboration published a letter of intent for a next-generation heavy-ion experiment ALICE 3\,\cite{ALICE:2022wwr}. Precision measurements of electromagnetic probes, i.e.\,virtual and real photons, are main pillars of the physics motivation for the new detector. The new all-silicon tracking system based monolithic active pixel sensor (MAPS) technology with a size of about 60\,$\rm m^{2}$ and a pseudo-rapidity coverage of $|\eta| < 4$ should provide the ultimate pointing resolution to see prompt thermal radiation from the partonic phase as a clear peak at small DCA$_{\rm ee}$ in the mass region $1.1 < \mll < 2.6$\,\GeVmass. An improvement by about a factor 10 compared to ALICE in Run 4 is expected for the pointing resolution. This will be achieved with a retractable super-light vertex detector, using the technology being currently developed for the ALICE innermost ITS layers in Run 4. Its main particularity is that it will be installed inside the beam pipe in a secondary vacuum. This way, the material budget in front of the silicon sensors can be reduced to the minimum, suppressing multiple scattering for low momentum particles, and the first tracking layer can be placed at 5\,mm to the beam axis (ALICE ITS Run 4: 19\,mm) during data taking. Since the aperture of the LHC beams at injection is about 16 mm, it has to be opened up when the beam is not yet stable. Therefore, the vertex detector will be installed on a movable so-called IRIS layout. Additional time-of-flight systems and a Cherenkov detector will enable the identification of low-momentum electrons. Finally, a muon system made of absorbers and tracking chambers placed outside of the magnet will complete the dilepton capabilities towards larger momenta. 
For electrons, the low \pt\,\,tracking and identification capability of the detector together with the excellent pointing resolution should offer unprecedented possibilities to perform precise differential measurements of thermal and pre-equilibrium dilepton production. In particular, mass dependent studies can be used to investigate the build-up of radial flow (via \ptll(\mll) or transverse mass $m_{\rm T,ll}(\mll)$), elliptic flow ($v_{\rm 2,ll}(\mll)$), and the evolution of the average effective temperature (slope of the excess mass spectrum $T_{\rm eff}(\mll)$) in the medium as a function of time. Such measurements would bring information on $\eta/s$\,\cite{Kasmaei:2018oag}, $\zeta/s$\,\cite{Vujanovic:2019yih} and $T$\,\cite{Rapp:2014hha,Churchill:2023zkk} in the interacting system and shed light on the long standing photon puzzle. While real photons at RHIC exhibit measured \pt-spectra in line with early emission from the hot phase of the system evolution\,\cite{PHENIX:2022rsx}, they show a large azimuthal asymmetry (elliptic flow)\,\cite{PHENIX:2011oxq,PHENIX:2015igl}, indicating significant photon emission at later times. These two observations do not fit easily into a single hydrodynamical description of the data\,\cite{Gale:2021emg}. At the LHC,
direct photon yield\,\cite{ALICE:2015xmh} and elliptic flow\,\cite{ALICE:2018dti} measurements in Pb--Pb collisions at $\sqrt{s_{\rm NN}} = 2.76$\,TeV do not show such puzzle within the experimental uncertainties, but the significance of the signal is still limited. Moreover, polarization measurements, which should be then in reach, are predicted to help to resolve the origin of the excess $\rm l^{+}l^{-}$ pairs at high masses (1.5-3\,\GeVmass) and determine if the lepton pairs are mainly produced thermally in the hot partonic phase, radiated from the partonic matter at a pre-equilibrium stage, or emitted from the annihilation of a quark and antiquark belonging to the two incoming nuclei (Drell-Yan process). While such information is mandatory to understand better the extracted effective temperatures from the excess mass spectra, polarization results as a function of mass are also expected to probe the equilibration time of the hot fireball connected to $\eta/s$ at the very early stage of the heavy-ion collision\,\cite{Coquet:2023wjk}.  
For masses below 1.2\,\GeVmass, detailed measurements of the $\rho$ spectral function are projected with a total uncertainty smaller than 8\,\% over a large mass range. This would allow a deep understanding of the chiral symmetry restoration mechanisms, including the $\rho-\rm a_{1}$ mixing process\,\cite{Hohler:2013eba}.
Finally, the potential for probing the electric conductivity of the hot medium with thermal radiation measurements at very low \mll\,\, and \ptll\,\, was also partially investigated and could be within reach\,\cite{Bailhache:2024mck}. \\


\section*{DISCLOSURE STATEMENT}
Both authors are members of the ALICE Collaboration. The authors are not aware of any other
affiliations, memberships, funding, or financial holdings that might be perceived as affecting the
objectivity of this review.

\section*{ACKNOWLEDGMENTS}
The authors thank M. Coquet, A. Drees, R. Esha, D. Gabor, T. Galatyuk, C. Gale, O. Garcia-
Montero, F. Geurts, R. Goes-Hirayama, J. Jung, J.-F. Paquet, R. Rapp, S. Scheid, S. Schlichting,
D. Sekihata, H. van Hees, and I. Vorobyev for insightful discussions.

\bibliographystyle{ar-style5.bst}

\begin{thebibliography}{96}
\expandafter\ifx\csname
natexlab\endcsname\relax\def\natexlab#1{#1}\fi






\bibitem{Bazavov:2011nk}
A.~Bazavov, T.~Bhattacharya, M.~Cheng, C.~DeTar, H.~T.~Ding, S.~Gottlieb, R.~Gupta, P.~Hegde, U.~M.~Heller and F.~Karsch, \textit{et al.}
Phys. Rev. D \textbf{85} (2012), 054503
doi:10.1103/PhysRevD.85.054503
[arXiv:1111.1710 [hep-lat]].

\bibitem{Borsanyi:2013bia}
S.~Borsanyi, Z.~Fodor, C.~Hoelbling, S.~D.~Katz, S.~Krieg and K.~K.~Szabo,
Phys. Lett. B \textbf{730} (2014), 99-104
doi:10.1016/j.physletb.2014.01.007
[arXiv:1309.5258 [hep-lat]].

\bibitem{HotQCD:2018pds}
A.~Bazavov \textit{et al.} [HotQCD],
Phys. Lett. B \textbf{795} (2019), 15-21
doi:10.1016/j.physletb.2019.05.013
[arXiv:1812.08235 [hep-lat]].

\bibitem{Borsanyi:2020fev}
S.~Borsanyi, Z.~Fodor, J.~N.~Guenther, R.~Kara, S.~D.~Katz, P.~Parotto, A.~Pasztor, C.~Ratti and K.~K.~Szabo,
Phys. Rev. Lett. \textbf{125} (2020) no.5, 052001
doi:10.1103/PhysRevLett.125.052001
[arXiv:2002.02821 [hep-lat]].

\bibitem{Lacey:2001va}
R.~A.~Lacey [PHENIX],
Nucl. Phys. A \textbf{698} (2002), 559-563
doi:10.1016/S0375-9474(01)01428-2
[arXiv:nucl-ex/0105003 [nucl-ex]].

\bibitem{Shuryak:2004cy}
E.~V.~Shuryak,
Nucl. Phys. A \textbf{750} (2005), 64-83
doi:10.1016/j.nuclphysa.2004.10.022
[arXiv:hep-ph/0405066 [hep-ph]].

\bibitem{Gyulassy:2004zy}
M.~Gyulassy and L.~McLerran,
Nucl. Phys. A \textbf{750} (2005), 30-63
doi:10.1016/j.nuclphysa.2004.10.034
[arXiv:nucl-th/0405013 [nucl-th]].





\bibitem{Rapp:2014hha}
R.~Rapp and H.~van Hees,
Phys. Lett. B \textbf{753} (2016), 586-590
doi:10.1016/j.physletb.2015.12.065
[arXiv:1411.4612 [hep-ph]].


\bibitem{Bilic:1997sh}
N.~Bilic and H.~Nikolic,
Eur. Phys. J. C \textbf{6} (1999), 515-523
doi:10.1007/s100529800923
[arXiv:hep-ph/9711513 [hep-ph]].

\bibitem{Dominguez:2012bs}
C.~A.~Dominguez, M.~Loewe and Y.~Zhang,
Phys. Rev. D \textbf{86} (2012) no.3, 034030
[erratum: Phys. Rev. D \textbf{90} (2014) no.3, 039903]
doi:10.1103/PhysRevD.86.034030
[arXiv:1205.3361 [hep-ph]].

\bibitem{Borsanyi:2010bp}
S.~Borsanyi \textit{et al.} [Wuppertal-Budapest],
JHEP \textbf{09} (2010), 073
doi:10.1007/JHEP09(2010)073
[arXiv:1005.3508 [hep-lat]].



\bibitem{Pisarski:1995xu}
R.~D.~Pisarski,
Phys. Rev. D \textbf{52} (1995), R3773-R3776
doi:10.1103/PhysRevD.52.R3773
[arXiv:hep-ph/9503328 [hep-ph]].

\bibitem{Rapp:1999us}
R.~Rapp and J.~Wambach,
Eur. Phys. J. A \textbf{6} (1999), 415-420
doi:10.1007/s100500050364
[arXiv:hep-ph/9907502 [hep-ph]].

\bibitem{vanHees:2007th}
H.~van Hees and R.~Rapp,
Nucl. Phys. A \textbf{806} (2008), 339-387
doi:10.1016/j.nuclphysa.2008.03.009
[arXiv:0711.3444 [hep-ph]].


\bibitem{Hohler:2013eba}
P.~M.~Hohler and R.~Rapp,
Phys. Lett. B \textbf{731} (2014), 103-109
doi:10.1016/j.physletb.2014.02.021
[arXiv:1311.2921 [hep-ph]].



\bibitem{Churchill:2023zkk}
J.~Churchill, L.~Du, C.~Gale, G.~Jackson and S.~Jeon,
Phys. Rev. Lett. \textbf{132} (2024) no.17, 172301
doi:10.1103/PhysRevLett.132.172301
[arXiv:2311.06951 [nucl-th]].

\bibitem{Churchill:2023vpt}
J.~Churchill, L.~Du, C.~Gale, G.~Jackson and S.~Jeon,
Phys. Rev. C \textbf{109} (2024) no.4, 044915
doi:10.1103/PhysRevC.109.044915
[arXiv:2311.06675 [nucl-th]].



\bibitem{Coquet:2021lca}
M.~Coquet, X.~Du, J.~Y.~Ollitrault, S.~Schlichting and M.~Winn,
Phys. Lett. B \textbf{821} (2021), 136626
doi:10.1016/j.physletb.2021.136626
[arXiv:2104.07622 [nucl-th]].

\bibitem{Coquet:2021gms}
M.~Coquet, X.~Du, J.~Y.~Ollitrault, S.~Schlichting and M.~Winn,
Nucl. Phys. A \textbf{1030} (2023), 122579
doi:10.1016/j.nuclphysa.2022.122579
[arXiv:2112.13876 [nucl-th]].

\bibitem{Coquet:2023wjk}
M.~Coquet, M.~Winn, X.~Du, J.~Y.~Ollitrault and S.~Schlichting,
Phys. Rev. Lett. \textbf{132} (2024) no.23, 232301
doi:10.1103/PhysRevLett.132.232301
[arXiv:2309.00555 [nucl-th]].

\bibitem{Vujanovic:2019yih}
G.~Vujanovic, J.~F.~Paquet, C.~Shen, G.~S.~Denicol, S.~Jeon, C.~Gale and U.~Heinz,
Phys. Rev. C \textbf{101} (2020), 044904
doi:10.1103/PhysRevC.101.044904
[arXiv:1903.05078 [nucl-th]].

\bibitem{Kasmaei:2018oag}
B.~S.~Kasmaei and M.~Strickland,
Phys. Rev. D \textbf{99} (2019) no.3, 034015
doi:10.1103/PhysRevD.99.034015
[arXiv:1811.07486 [hep-ph]].

\bibitem{Huang:2015oca}
X.~G.~Huang,
Rept. Prog. Phys. \textbf{79} (2016) no.7, 076302
doi:10.1088/0034-4885/79/7/076302
[arXiv:1509.04073 [nucl-th]].

\bibitem{Tuchin:2015oka}
K.~Tuchin,
Phys. Rev. C \textbf{93} (2016) no.1, 014905
doi:10.1103/PhysRevC.93.014905
[arXiv:1508.06925 [hep-ph]].

\bibitem{Fotakis:2021diq}
J.~A.~Fotakis, O.~Soloveva, C.~Greiner, O.~Kaczmarek and E.~Bratkovskaya,
Phys. Rev. D \textbf{104} (2021) no.3, 034014
doi:10.1103/PhysRevD.104.034014
[arXiv:2102.08140 [hep-ph]].


\bibitem{Ding:2016hua}
H.~T.~Ding, O.~Kaczmarek and F.~Meyer,
Phys. Rev. D \textbf{94} (2016) no.3, 034504
doi:10.1103/PhysRevD.94.034504
[arXiv:1604.06712 [hep-lat]].

\bibitem{Moore:2006qn}
G.~D.~Moore and J.~M.~Robert,
[arXiv:hep-ph/0607172 [hep-ph]].

\bibitem{Floerchinger:2021xhb}
S.~Floerchinger, C.~Gebhardt and K.~Reygers,
Phys. Lett. B \textbf{837} (2023), 137647
doi:10.1016/j.physletb.2022.137647
[arXiv:2112.12497 [nucl-th]].

\bibitem{Rapp:2024grb}
R.~Rapp,
[arXiv:2406.14656 [hep-ph]].

\bibitem{PHENIX:2009gyd}
A.~Adare \textit{et al.} [PHENIX],
Phys. Rev. C \textbf{81} (2010), 034911
doi:10.1103/PhysRevC.81.034911
[arXiv:0912.0244 [nucl-ex]].


\bibitem{Breit:1934zz}
G.~Breit and J.~A.~Wheeler,
Phys. Rev. \textbf{46} (1934) no.12, 1087-1091
doi:10.1103/PhysRev.46.1087

\bibitem{STAR:2018ldd}
J.~Adam \textit{et al.} [STAR],
Phys. Rev. Lett. \textbf{121} (2018) no.13, 132301
doi:10.1103/PhysRevLett.121.132301
[arXiv:1806.02295 [hep-ex]].

\bibitem{ATLAS:2018pfw}
M.~Aaboud \textit{et al.} [ATLAS],
Phys. Rev. Lett. \textbf{121} (2018) no.21, 212301
doi:10.1103/PhysRevLett.121.212301
[arXiv:1806.08708 [nucl-ex]].

\bibitem{ATLAS:2022yad}
G.~Aad \textit{et al.} [ATLAS],
Phys. Rev. C \textbf{107} (2023) no.5, 054907
doi:10.1103/PhysRevC.107.054907
[arXiv:2206.12594 [nucl-ex]].

\bibitem{ALICE:2022hvk}
S.~Acharya \textit{et al.} [ALICE],
JHEP \textbf{06} (2023), 024
doi:10.1007/JHEP06(2023)024
[arXiv:2204.11732 [nucl-ex]].

\bibitem{Brandenburg:2021lnj}
J.~D.~Brandenburg, W.~Zha and Z.~Xu,
Eur. Phys. J. A \textbf{57} (2021) no.10, 299
doi:10.1140/epja/s10050-021-00595-5
[arXiv:2103.16623 [hep-ph]].




\bibitem{NA38:1993pef}
M.~C.~Abreu \textit{et al.} [NA38],
Nucl. Phys. A \textbf{566} (1994), 77C-85C
doi:10.1016/0375-9474(94)90611-4

\bibitem{HELIOS3:a}
M.~Masera \textit{et al.} [HELIOS/3],
Nucl. Phys. A \textbf{590} (1995), 93-102
doi:10.1016/0375-9474(95)00228-S

\bibitem{HELIOS3:1998xeb}
A.~L.~S.~Angelis \textit{et al.} [HELIOS/3],
Eur. Phys. J. C \textbf{13} (2000), 433-452
doi:10.1007/s100520050707

\bibitem{CERES:1995vll}
G.~Agakichiev \textit{et al.} [CERES],
Phys. Rev. Lett. \textbf{75} (1995), 1272-1275
doi:10.1103/PhysRevLett.75.1272

\bibitem{CERESNA45:1997tgc}
G.~Agakichiev \textit{et al.} [CERES/NA45],
Phys. Lett. B \textbf{422} (1998), 405-412
doi:10.1016/S0370-2693(98)00083-5
[arXiv:nucl-ex/9712008 [nucl-ex]].

\bibitem{CERESNA45:2002gnc}
D.~Adamova \textit{et al.} [CERES/NA45],
Phys. Rev. Lett. \textbf{91} (2003), 042301
doi:10.1103/PhysRevLett.91.042301
[arXiv:nucl-ex/0209024 [nucl-ex]].

\bibitem{CERES:2006wcq}
D.~Adamova \textit{et al.} [CERES],
Phys. Lett. B \textbf{666} (2008), 425-429
doi:10.1016/j.physletb.2008.07.104
[arXiv:nucl-ex/0611022 [nucl-ex]].

\bibitem{NA60:2006ymb}
R.~Arnaldi \textit{et al.} [NA60],
Phys. Rev. Lett. \textbf{96} (2006), 162302
doi:10.1103/PhysRevLett.96.162302
[arXiv:nucl-ex/0605007 [nucl-ex]].

\bibitem{NA60:2008dcb}
R.~Arnaldi \textit{et al.} [NA60],
Eur. Phys. J. C \textbf{59} (2009), 607-623
doi:10.1140/epjc/s10052-008-0857-2
[arXiv:0810.3204 [nucl-ex]].

\bibitem{NA60:2008ctj}
R.~Arnaldi \textit{et al.} [NA60],
Eur. Phys. J. C \textbf{61} (2009), 711-720
doi:10.1140/epjc/s10052-009-0878-5
[arXiv:0812.3053 [nucl-ex]].

\bibitem{Specht:2010xu}
H.~J.~Specht [NA60],
AIP Conf. Proc. \textbf{1322} (2010) no.1, 1-10
doi:10.1063/1.3541982
[arXiv:1011.0615 [nucl-ex]].

\bibitem{Brown:1995qt}
G.~E.~Brown and M.~Rho,
Phys. Rept. \textbf{269} (1996), 333-380
doi:10.1016/0370-1573(95)00067-4
[arXiv:hep-ph/9504250 [hep-ph]].

\bibitem{Rapp:1999ej}
R.~Rapp and J.~Wambach,
Adv. Nucl. Phys. \textbf{25} (2000), 1
doi:10.1007/0-306-47101-9\_1
[arXiv:hep-ph/9909229 [hep-ph]].

\bibitem{Eletsky:2001bb}
V.~L.~Eletsky, M.~Belkacem, P.~J.~Ellis and J.~I.~Kapusta,
Phys. Rev. C \textbf{64} (2001), 035202
doi:10.1103/PhysRevC.64.035202
[arXiv:nucl-th/0104029 [nucl-th]].

\bibitem{Rapp:2009yu}
R.~Rapp, J.~Wambach and H.~van Hees,
Landolt-Bornstein \textbf{23} (2010), 134
doi:10.1007/978-3-642-01539-7\_6
[arXiv:0901.3289 [hep-ph]].

\bibitem{Xu:2011tz}
H.~j.~Xu, H.~f.~Chen, X.~Dong, Q.~Wang and Y.~f.~Zhang,
Phys. Rev. C \textbf{85} (2012), 024906
doi:10.1103/PhysRevC.85.024906
[arXiv:1110.4825 [nucl-th]].

\bibitem{Vujanovic:2013jpa}
G.~Vujanovic, C.~Young, B.~Schenke, R.~Rapp, S.~Jeon and C.~Gale,
Phys. Rev. C \textbf{89} (2014) no.3, 034904
doi:10.1103/PhysRevC.89.034904
[arXiv:1312.0676 [nucl-th]].

\bibitem{Rapp:2013nxa}
R.~Rapp,
Adv. High Energy Phys. \textbf{2013} (2013), 148253
doi:10.1155/2013/148253
[arXiv:1304.2309 [hep-ph]].

\bibitem{Cassing:2009vt}
W.~Cassing and E.~L.~Bratkovskaya,
Nucl. Phys. A \textbf{831} (2009), 215-242
doi:10.1016/j.nuclphysa.2009.09.007
[arXiv:0907.5331 [nucl-th]].

\bibitem{Bratkovskaya:2011wp}
E.~L.~Bratkovskaya, W.~Cassing, V.~P.~Konchakovski and O.~Linnyk,
Nucl. Phys. A \textbf{856} (2011), 162-182
doi:10.1016/j.nuclphysa.2011.03.003
[arXiv:1101.5793 [nucl-th]].

\bibitem{Song:2018xca}
T.~Song, W.~Cassing, P.~Moreau and E.~Bratkovskaya,
Phys. Rev. C \textbf{97} (2018) no.6, 064907
doi:10.1103/PhysRevC.97.064907
[arXiv:1803.02698 [nucl-th]].

\bibitem{Giacalone:2019ldn}
G.~Giacalone, A.~Mazeliauskas and S.~Schlichting,
Phys. Rev. Lett. \textbf{123} (2019) no.26, 262301
doi:10.1103/PhysRevLett.123.262301
[arXiv:1908.02866 [hep-ph]].

\bibitem{Kurkela:2018wud}
A.~Kurkela, A.~Mazeliauskas, J.~F.~Paquet, S.~Schlichting and D.~Teaney,
Phys. Rev. Lett. \textbf{122} (2019) no.12, 122302
doi:10.1103/PhysRevLett.122.122302
[arXiv:1805.01604 [hep-ph]].

\bibitem{Kurkela:2018xxd}
A.~Kurkela and A.~Mazeliauskas,
Phys. Rev. Lett. \textbf{122} (2019), 142301
doi:10.1103/PhysRevLett.122.142301
[arXiv:1811.03040 [hep-ph]].

\bibitem{Du:2020zqg}
X.~Du and S.~Schlichting,
Phys. Rev. Lett. \textbf{127} (2021) no.12, 122301
doi:10.1103/PhysRevLett.127.122301
[arXiv:2012.09068 [hep-ph]].

\bibitem{Garcia-Montero:2024msw}
O.~Garcia-Montero, A.~Mazeliauskas, P.~Plaschke and S.~Schlichting,
EPJ Web Conf. \textbf{296} (2024), 07003
doi:10.1051/epjconf/202429607003
[arXiv:2404.02861 [hep-ph]].


\bibitem{STAR:2015tnn}
L.~Adamczyk \textit{et al.} [STAR],
Phys. Rev. C \textbf{92} (2015) no.2, 024912
doi:10.1103/PhysRevC.92.024912
[arXiv:1504.01317 [hep-ex]].

\bibitem{PHENIX:2015vek}
A.~Adare \textit{et al.} [PHENIX],
Phys. Rev. C \textbf{93} (2016) no.1, 014904
doi:10.1103/PhysRevC.93.014904
[arXiv:1509.04667 [nucl-ex]].

\bibitem{ALICE:2018ael}
S.~Acharya \textit{et al.} [ALICE],
Phys. Rev. C \textbf{99} (2019) no.2, 024002
doi:10.1103/PhysRevC.99.024002
[arXiv:1807.00923 [nucl-ex]].

\bibitem{ALICE:2023jef}
S.~Acharya \textit{et al.} [ALICE],
Phys. Rev. C \textbf{112} (2025) no.5, 054906
doi:10.1103/xl6m-vbqk
[arXiv:2308.16704 [nucl-ex]].




\bibitem{Linnyk:2011vx}
O.~Linnyk, W.~Cassing, J.~Manninen, E.~L.~Bratkovskaya and C.~M.~Ko,
Phys. Rev. C \textbf{85} (2012), 024910
doi:10.1103/PhysRevC.85.024910
[arXiv:1111.2975 [nucl-th]].


\bibitem{STAR:2013pwb}
L.~Adamczyk \textit{et al.} [STAR],
Phys. Rev. Lett. \textbf{113} (2014) no.2, 022301
doi:10.1103/PhysRevLett.113.022301
[arXiv:1312.7397 [hep-ex]].

\bibitem{PHENIX:2014edx}
A.~Adare \textit{et al.} [PHENIX],
Phys. Rev. C \textbf{91} (2015) no.1, 014907
doi:10.1103/PhysRevC.91.014907
[arXiv:1405.4004 [nucl-ex]].


\bibitem{PHENIX:2017ztp}
A.~Adare \textit{et al.} [PHENIX],
Phys. Rev. C \textbf{96} (2017) no.2, 024907
doi:10.1103/PhysRevC.96.024907
[arXiv:1702.01084 [nucl-ex]].

\bibitem{PHENIX:2006iih}
A.~Adare \textit{et al.} [PHENIX],
Phys. Rev. Lett. \textbf{98} (2007), 172301
doi:10.1103/PhysRevLett.98.172301
[arXiv:nucl-ex/0611018 [nucl-ex]].


\bibitem{STAR:2014wif}
L.~Adamczyk \textit{et al.} [STAR],
Phys. Rev. Lett. \textbf{113} (2014) no.14, 142301
[erratum: Phys. Rev. Lett. \textbf{121} (2018) no.22, 229901]
doi:10.1103/PhysRevLett.113.142301
[arXiv:1404.6185 [nucl-ex]].



\bibitem{STAR:2015zal}
L.~Adamczyk \textit{et al.} [STAR],
Phys. Lett. B \textbf{750} (2015), 64-71
doi:10.1016/j.physletb.2015.08.044
[arXiv:1501.05341 [hep-ex]].

\bibitem{STAR:2023wta}
M.~I.~Abdulhamid \textit{et al.} [STAR],
Phys. Rev. C \textbf{107} (2023) no.6, L061901
doi:10.1103/PhysRevC.107.L061901

\bibitem{Han:2024nzr}
Y.~Han [STAR],
EPJ Web Conf. \textbf{296} (2024), 07004
doi:10.1051/epjconf/202429607004

\bibitem{sQM2024starprel}
Z.~Wang [STAR],
Presentation at the 21st International Conference on Strangeness in Quark Matter (SQM 2024), Jun. 3–7., https://indico.in2p3.fr/event/29792/contributions/137176/ (2024)

\bibitem{HP2024starprel}
C.~Jin [STAR],
EPJ Web Conf. \textbf{339} (2025), 05005
doi:10.1051/epjconf/202533905005
[arXiv:2505.06361 [nucl-ex]].


\bibitem{STAR:2024bpc}
B.~E.~Aboona \textit{et al.} [STAR],
Nature Commun. \textbf{16} (2025) no.1, 9098
doi:10.1038/s41467-025-63216-5
[arXiv:2402.01998 [nucl-ex]].

\bibitem{HP2024starprelbis}
J.~Luo [STAR],
Presentation at the 12th International Conference on Hard and Electro-
magnetic Probes of High-Energy Nuclear Collisions (Hard Probes 2024), Sept. 22–27. https://indico.
cern.ch/event/1339555/contributions/6040892/ (2024)

\bibitem{STAR:2017sal}
L.~Adamczyk \textit{et al.} [STAR],
Phys. Rev. C \textbf{96} (2017) no.4, 044904
doi:10.1103/PhysRevC.96.044904
[arXiv:1701.07065 [nucl-ex]].

\bibitem{Braun-Munzinger:2003htr}
P.~Braun-Munzinger, J.~Stachel and C.~Wetterich,
Phys. Lett. B \textbf{596} (2004), 61-69
doi:10.1016/j.physletb.2004.05.081
[arXiv:nucl-th/0311005 [nucl-th]].

\bibitem{CBM:2016kpk}
T.~Ablyazimov \textit{et al.} [CBM],
Eur. Phys. J. A \textbf{53} (2017) no.3, 60
doi:10.1140/epja/i2017-12248-y
[arXiv:1607.01487 [nucl-ex]].

\bibitem{NA60:2022sze}
C.~Ahdida \textit{et al.} [NA60+],
[arXiv:2212.14452 [nucl-ex]].

\bibitem{HADES:2019auv}
J.~Adamczewski-Musch \textit{et al.} [HADES],
Nature Phys. \textbf{15} (2019) no.10, 1040-1045
doi:10.1038/s41567-019-0583-8


%
%

\bibitem{LHCb:2017trq}
R.~Aaij \textit{et al.} [LHCb],
Phys. Rev. Lett. \textbf{120} (2018) no.6, 061801
doi:10.1103/PhysRevLett.120.061801
[arXiv:1710.02867 [hep-ex]].

\bibitem{LHCb:2018roe}
R.~Aaij \textit{et al.} [LHCb],
[arXiv:1808.08865 [hep-ex]].

\bibitem{ALICE:2020mfy}
S.~Acharya \textit{et al.} [ALICE],
Phys. Rev. C \textbf{102} (2020) no.5, 055204
doi:10.1103/PhysRevC.102.055204
[arXiv:2005.11995 [nucl-ex]].

\bibitem{ALICE:2019nuy}
S.~Acharya \textit{et al.} [ALICE],
Phys. Lett. B \textbf{804} (2020), 135377
doi:10.1016/j.physletb.2020.135377
[arXiv:1910.09110 [nucl-ex]].

\bibitem{Eskola:2009uj}
K.~J.~Eskola, H.~Paukkunen and C.~A.~Salgado,
JHEP \textbf{04} (2009), 065
doi:10.1088/1126-6708/2009/04/065
[arXiv:0902.4154 [hep-ph]].


%
%

\bibitem{Arduini:2024kdp}
G.~Arduini, V.~Baglin, H.~Bartosik, L.~Bottura, C.~Bracco, B.~Bradu, G.~Bregliozzi, K.~Brodzinski, R.~Bruce and M.~Calviani, \textit{et al.}
JINST \textbf{19} (2024) no.05, P05061
doi:10.1088/1748-0221/19/05/P05061


\bibitem{ALICE:2012dtf}
B.~Abelev \textit{et al.} [ALICE],
J. Phys. G \textbf{41} (2014), 087001
doi:10.1088/0954-3899/41/8/087001

\bibitem{ALICE:2023udb}
S.~Acharya \textit{et al.} [ALICE],
JINST \textbf{19} (2024) no.05, P05062
doi:10.1088/1748-0221/19/05/P05062
[arXiv:2302.01238 [physics.ins-det]].

\bibitem{Lippmann:2014lay}
Z. Ahammed \textit{et al.} [ALICE],
CERN-LHCC-2013-020.

\bibitem{TheALICECollaboration:2015xke}
J. Adam \textit{et al.} [ALICE], 
CERN-LHCC-2015-002.

\bibitem{ALICETPC:2020ann}
J.~Adolfsson \textit{et al.} [ALICE TPC],
JINST \textbf{16} (2021) no.03, P03022
doi:10.1088/1748-0221/16/03/P03022
[arXiv:2012.09518 [physics.ins-det]].

\bibitem{ALICE:2013nwm}
B.~Abelev \textit{et al.} [ALICE],
J. Phys. G \textbf{41} (2014), 087002
doi:10.1088/0954-3899/41/8/087002

\bibitem{ITS3}
S.~Acharya \textit{et al.} [ALICE],
CERN-LHCC-2019-018.

\bibitem{ITS3bis}
S.~Acharya \textit{et al.} [ALICE],
CERN-LHCC-2024-003.

\bibitem{Citron:2018lsq}
Z.~Citron, A.~Dainese, J.~F.~Grosse-Oetringhaus, J.~M.~Jowett, Y.~J.~Lee, U.~A.~Wiedemann, M.~Winn, A.~Andronic, F.~Bellini and E.~Bruna, \textit{et al.}
CERN Yellow Rep. Monogr. \textbf{7} (2019), 1159-1410
doi:10.23731/CYRM-2019-007.1159
[arXiv:1812.06772 [hep-ph]].

\bibitem{ALICE:2018fvj}
S.~Acharya \textit{et al.} [ALICE],
JHEP \textbf{09} (2018), 064
doi:10.1007/JHEP09(2018)064
[arXiv:1805.04391 [hep-ex]].

\bibitem{mft}
J. Adam \textit{et al.} [ALICE], 
CERN-LHCC-2015-001.

\bibitem{JeromeJungHardProbes2024}
J.~Jung [ALICE],
[arXiv:2505.03669 [nucl-ex]].

%
%

\bibitem{LHCb:2019gvd}
R.~Aaij \textit{et al.} [LHCb],
JINST \textbf{14} (2019), P11023
doi:10.1088/1748-0221/14/11/P11023
[arXiv:1909.02957 [hep-ex]].

\bibitem{ALICE:2022wwr}
S.~Acharya \textit{et al.} [ALICE],
[arXiv:2211.02491 [physics.ins-det]].

\bibitem{PHENIX:2022rsx}
N.~J.~Abdulameer \textit{et al.} [PHENIX],
Phys. Rev. C \textbf{109} (2024) no.4, 044912
doi:10.1103/PhysRevC.109.044912
[arXiv:2203.17187 [nucl-ex]].

\bibitem{PHENIX:2011oxq}
A.~Adare \textit{et al.} [PHENIX],
Phys. Rev. Lett. \textbf{109} (2012), 122302
doi:10.1103/PhysRevLett.109.122302
[arXiv:1105.4126 [nucl-ex]].

\bibitem{PHENIX:2015igl}
A.~Adare \textit{et al.} [PHENIX],
Phys. Rev. C \textbf{94} (2016) no.6, 064901
doi:10.1103/PhysRevC.94.064901
[arXiv:1509.07758 [nucl-ex]].


\bibitem{Gale:2021emg}
C.~Gale, J.~F.~Paquet, B.~Schenke and C.~Shen,
Phys. Rev. C \textbf{105} (2022) no.1, 014909
doi:10.1103/PhysRevC.105.014909
[arXiv:2106.11216 [nucl-th]].

\bibitem{ALICE:2015xmh}
J.~Adam \textit{et al.} [ALICE],
Phys. Lett. B \textbf{754} (2016), 235-248
doi:10.1016/j.physletb.2016.01.020
[arXiv:1509.07324 [nucl-ex]].

\bibitem{ALICE:2018dti}
S.~Acharya \textit{et al.} [ALICE],
Phys. Lett. B \textbf{789} (2019), 308-322
doi:10.1016/j.physletb.2018.11.039
[arXiv:1805.04403 [nucl-ex]].

\bibitem{Bailhache:2024mck}
R.~Bailhache, D.~Bonocore, P.~Braun-Munzinger, X.~Feal, S.~Floerchinger, J.~Klein, K.~K{\"o}hler, P.~Lebiedowicz, C.~M.~Peter and R.~Rapp, \textit{et al.}
Phys. Rept. \textbf{1097} (2024), 1-40
doi:10.1016/j.physrep.2024.10.002
[arXiv:2406.17959 [nucl-ex]].




\end{thebibliography}

\end{document}